\documentclass[preprint]{aastex}

\newcommand{\be}{\begin{equation}}
\newcommand{\ee}{\end{equation}}
\newcommand{\bea}{\begin{eqnarray}}
\newcommand{\eea}{\end{eqnarray}}
\newcommand{\var}{\mathrm{var}}
\newcommand{\go}{\mathrel{\raise.3ex\hbox{$>$}\mkern-14mu\lower0.6ex\hbox{$\sim$}}}
\newcommand{\lo}{\mathrel{\raise.3ex\hbox{$<$}\mkern-14mu\lower0.6ex\hbox{$\sim$}}}

\slugcomment{ApJ, in press}
\shorttitle{MCMC for Extrasolar Planets}
\shortauthors{Ford}
\begin{document}

\title{Improving the Efficiency of Markov Chain Monte Carlo for 
Analyzing the Orbits of Extrasolar Planets}

\author{Eric B.\ Ford}

\affil{Astronomy Department, 
	601 Campbell Hall, 
	University of California at Berkeley, 
	Berkeley, CA 94720-3411, USA}
\affil{Department of Astrophysical Sciences, 
	Princeton University, 
	Peyton Hall, 
	Princeton, NJ 08544-1001, USA}
\email{eford@astron.berkeley.edu}

\begin{abstract}
Precise radial velocity measurements have led to the discovery of
$\sim170$ extrasolar planetary systems.  Understanding the
uncertainties in the orbital solutions will become increasingly
important as the discovery space for extrasolar planets shifts to
planets with smaller masses and longer orbital periods.  The method of
Markov chain Monte Carlo (MCMC) provides a rigorous method for
quantifying the uncertainties in orbital parameters in a Bayesian
framework (Ford 2005a).  The main practical challenge for the general
application of MCMC is the need to construct Markov chains which
quickly converge.  The rate of convergence is very sensitive to the
choice of the candidate transition probability distribution function
(CTPDF).  Here we explain one simple method for generating alternative
CTPDFs which can significantly speed convergence by one to three
orders of magnitude.  We have numerically tested dozens of CTPDFs with
simulated radial velocity data sets to identify those which perform
well for different types of orbits and suggest a set of CTPDFs for
general application.  Additionally, we introduce other refinements to
the MCMC algorithm for radial velocity planets, including an improved
treatment of the uncertainties in the radial velocity observations, an
algorithm for automatically choosing step sizes, an algorithm for
automatically determining reasonable stopping times, and the use of
importance sampling for including the dynamical evolution of multiple
planet systems.  Together, these improvements make it practical to
apply MCMC to multiple planet systems.  We demonstrate the
improvements in efficiency by analyzing a variety of extrasolar
planetary systems.
\end{abstract}

\keywords{Subject headings: planetary systems -- methods: statistical
-- techniques: radial velocities}

\section{Introduction}

Recent detections of planets around other stars have spurred a wide
range of research on planet formation and planetary system evolution.
The future of radial velocity planet searches promises to be exciting.
Ongoing large surveys including a broad array of nearby main-sequence
stars will continue to increase the number of known extrasolar
planets.  The challenges of accurately determining orbital parameters
will become even more important for three reasons.  First, continued
monitoring with the radial velocity technique will permit the
detection of planets with smaller masses.  Since the lowest mass
planet will always be near the threshold of detection, they will
typically have relatively low signal-to-noise ratios and hence
relatively large uncertainties in model parameters.  Second, the
increasing time span of precision observations will permit the
discovery of planets with larger orbital periods.  Unfortunately,
there can be large degeneracies in the orbital parameters for planets
with orbital periods comparable to the duration of observations.
Finally, the increasing precision and number of radial velocity
observations is likely to reveal more multiple planet systems.  The
large number of model parameters needed to model multiple planet
systems can lead to degeneracies and some orbital parameters being
poorly constrained.  All of these trends imply that it will become
increasingly important to understand the uncertainties in orbital
elements and other parameters derived from such observations. Thus, we
must use the best possible statistical tools to analyze radial
velocity data.

\subsection{Introduction to Bayesian Inference}

To quantitatively analyze the available observational constraints, we
employ the techniques of Bayesian inference.  The essential equations
of Bayesian inference can be easily derived from the basic axioms of
probability theory.  We start with a joint probability distribution,
$p(x,y)$, ($x$ and $y$ may be scalars or vectors of several
variables).  We then construct a marginalized probability distribution
for $x$ by integrating over $y$, $p(x) = \int p(x,y)\, dy$.  We can
then write the joint probability distribution as a product of the
marginalized probability distribution and a conditional probability
distribution, $p(x,y) = p(x) p(y|x)$.  Then, Bayes' theorem could be
written as
\be
p(y|x) = \frac{p(x,y)}{p(x)} = \frac{p(y) p(x|y)}{\int p(y) p(x|y) \, dy }.
\ee
The real insight in Bayes' theorem is to identify $x$ with a set of
observational data ($\vec{d}$) and $y$ with a set of model parameters
($\vec{\theta}$) that are not directly observed.  By treating both the
observation and the model parameters as random variables, Bayesian
inference is able to address statistical questions in a mathematically
rigorous fashion.  The joint probability, $p(\vec{d}, \vec{\theta})$,
can be expressed as the product of the likelihood ($p(\vec{d} |
\vec{\theta})$, the probability of the observables given the model
parameters), and a prior probability distribution function
($p(\vec{\theta})$) which is based on previous knowledge of the model
parameters.  Bayes's theorem allows one to compute a posterior
probability density function, $p(\vec{\theta} | \vec{d})$, which
incorporates the knowledge gained by the observations $\vec{d}$.  That
is
\bea
p(\vec{\theta}| \vec{d}, \mathcal{M} )  &
= &\frac{ p( \vec{d}, \vec{\theta} | \mathcal{M} ) }{ p(\vec{d | \mathcal{M} })}
= \frac{ p( \vec{d}, \vec{\theta} | \mathcal{M} ) }{ \int p( \vec{d}, \vec{\theta} | \mathcal{M} ) p( \vec{\theta} | \mathcal{M} ) \,d\vec{\theta} } \nonumber \\ 
& = & \frac{ p( \vec{\theta} | \mathcal{M} ) p(\vec{d} | \vec{\theta}, \mathcal{M} ) }{ \int p( \vec{\theta} | \mathcal{M} ) p( \vec{d}| \vec{\theta}, \mathcal{M} ) \,d\vec{\theta} },
\label{BayesEqn}
\eea
where we have now explicitly added the fact that each of the
probability distributions is conditioned on the assumption of a
certain model, $\mathcal{M}$, that includes the meaning of the model
parameters, $\vec{\theta}$, and their relationship to the
observational data, $\vec{d}$.  Table \ref{Tab1} provides a summary of
the symbols which appear in multiple sections of this paper.

\subsubsection{Advantages of Bayesian Inference}
 
Bayesian inference has several advantages over classical statistical
techniques.  First, the Bayesian framework provides a rigorous
mathematical foundation for making inferences about the model
parameters.  Since all Bayesian inferences are based on the posterior
probability distributions, there is a rigorous basis for quantifying
uncertainties in model parameters, unlike frequentist methods which
typically result in point estimates.  Frequentist methods are often
combined with resampling techniques such as bootstrap to estimate
uncertainties in model parameters.  However, these techniques rely on
fictitious observations that the experimenter believes could have been
made, while the Bayesian posterior probability distribution depends
only on the observations that were actually made (and the prior
knowledge of the model parameters).  While the posterior probability
distribution is often abbreviated as simply the posterior, it is
important to remember that it is a true probability distribution for
the model parameters, unlike the results of resampling techniques such
as bootstrap.  Because of these theoretical advantages, Bayesian
inference typically produces more accurate estimates of the
uncertainty in model parameters, as shown using actual radial velocity
observations of extrasolar planet in Ford (2005a, Paper I).  For
further discussion of alternative techniques see Paper I, \S 3 \& 4.

The Bayesian framework also has several practical advantages.  Because
inference is based on probability distributions, it is straightforward
to incorporate a variety of different types of information and
observations.  Further, hierarchical Bayesian models can naturally
accommodate uncertainties in models as well as in observations.
Finally, the Bayesian framework provides a natural basis for making
predictions about future observations.  These predictions can even be
used to improve the efficiency of future experiments or observations
(Loredo 2003; Ford 2005b).

The most commonly criticized aspect of Bayesian inference is the
necessity of specifying prior probability distributions for the model
parameters.  In many cases, the observational data provides such a
strong constraint on the model parameters that the effect of the prior
is negligible.  However, in cases where the observations provide only
very limited constraints, the posterior distribution can be
significantly influenced by the choice of prior.  A complete Bayesian
analysis should include checking the sensitivity of any conclusions to
the choice of prior.  It should be noted that this criticism is not
unique to Bayesian inference.  Similar assumptions are being made
implicitly in frequentist analyses, as the choice of model parameters
can also significantly influence the results.  For a striking example
of practical differences for orbit determination of binary stars, see
Pourbaix (2002).  Indeed, the necessity to explicitly state what
priors are being chosen can be viewed as a strength of the Bayesian
method.

Given all the advantages of Bayesian inference, one might wonder why
it is not the standard practice for data analysis in the physical
sciences.  Unfortunately, the lower integral in Eqn.\ \ref{BayesEqn}
can be extremely difficult to compute, particularly when
$\vec{\theta}$ has a large number of dimensions.  The computational
burden has long limited Bayesian methods to a small number of relatively
simple problems.  Modern computers permit the application of Bayesian
inference to an growing number of increasingly complex problem.  Even
today, the brute force evaluation of the integrals is impractical for
many real world problems.  Therefore, it is necessary to develop
efficient algorithms for evaluating the necessary integrals.  The
method of Markov chain Monte Carlo is quite general and it has become
an increasingly common tool for performing the necessary Bayesian
integrals in recent years (e.g., Paper I and reference therein).

\subsection{Introduction to Markov chain Monte Carlo}

Paper I introduced Bayesian analysis for constraining the orbital
parameters of extrasolar planets with radial velocity observations and
presented an algorithm to perform the necessary integrations based on
Markov chain Monte Carlo (MCMC) simulation.  This algorithm can accurately
characterize the posterior probability distribution function for
orbital parameters based on radial velocity observations.

Paper I described how to construct a Markov chain (i.e. sequence) of
states (i.e. sets of parameter values, $\vec{\theta}_i$) which are
sampled from the posterior probability function.  Such a chain can be
calculated by specifying an initial set of parameter values,
$\vec{\theta}_0$, and a transition probability, $p(\vec{\theta}_{i+1}
| \vec{\theta}_i, \mathcal{M})$.  To guarantee that the Markov chain
will converge to the posterior probability distribution, the Markov
chain must be aperiodic, irreducible (i.e., it must be possible for
the chain to reach every state with non-zero probability from any
other state with non-zero probability), and reversible, that is,
\begin{equation}
p(\vec{\theta}| \vec{d}, \mathcal{M} ) p(\vec{\theta}|\vec{\theta}', \mathcal{M}) = p(\vec{\theta}'| \vec{d}, \mathcal{M}) p(\vec{\theta}'|\vec{\theta}, \mathcal{M}).
\end{equation}
It is possible to construct a reversible transition probability,
$p(\vec{\theta}'|\vec{\theta}, \mathcal{M} )$, from a non-reversible
candidate transition probability distribution function (CTPDF),
$q(\vec{\theta}'|\vec{\theta}, \mathcal{M})$, using the
Metropolis-Hastings (MH) algorithm.  The MH algorithm involves the
generation of a trial state ($\vec{\theta}'$) according to the CTPDF,
$q(\vec{\theta}'|\vec{\theta}, \mathcal{M})$, and randomly accepting
the trial as the next state or rejecting the trial state in favor of
the current state.  The MH algorithm specifies an acceptance
probability
\bea
\alpha(\vec{\theta}'|\vec{\theta}, \mathcal{M}) & = & \min \left\{ \frac{  p(\vec{\theta}'| \vec{d}, \mathcal{M})
q(\vec{\theta}|\vec{\theta}', \mathcal{M})}{p(\vec{\theta}| \vec{d}, \mathcal{M}) q(\vec{\theta}'|\vec{\theta}, \mathcal{M}) }, 1 \right\} \\
& = & \min \left\{ \frac{  p(\vec{d} | \vec{\theta}', \mathcal{M})
q(\vec{\theta}|\vec{\theta}', \mathcal{M})}{p(\vec{d}|\vec{\theta}, \mathcal{M}) q(\vec{\theta}'|\vec{\theta}, \mathcal{M}) }, 1 \right\}.
\label{AlphaEqn}
\eea
When using this acceptance probability, the transition probability
\begin{equation}
p(\vec{\theta}'|\vec{\theta}, \mathcal{M}) = q(\vec{\theta}'|\vec{\theta}, \mathcal{M}) \alpha(\vec{\theta}'|\vec{\theta}, \mathcal{M})
\end{equation}
is guaranteed to be reversible and irreducible, provided only that
$q(\vec{\theta}'|\vec{\theta}, \mathcal{M})$ allows transitions to all
$\vec{\theta}'$ for which $p(\vec{\theta}' | \vec{d}, \mathcal{M})$ is
non-zero.  Note that the MH algorithm does not require that the
normalization of $p(\vec{\theta}| \vec{d}, \mathcal{M})$ be known.

While the above algorithm guarantees that the Markov chain will
converge to $p(\vec{\theta}| \vec{d}, \mathcal{M})$, it does not
specify when the chain will achieve convergence.  The choice of
$q(\vec{\theta}'|\vec{\theta}, \mathcal{M})$ can have a dramatic
effect on the rate of convergence of the Markov chain.  Poor choices
can lead to extremely inefficient sampling and hence slow convergence.
The most efficient choice for $q(\vec{\theta}'|\vec{\theta},
\mathcal{M})$ would be $p(\vec{\theta}' | \vec{d}, \mathcal{M} )$, the
posterior probability distribution itself.  However, this is rarely
possible, since the whole purpose of the Markov chain is to calculate
the posterior distribution.

Paper I presented a practical algorithm based on the 
Metroplis-Hastings algorithm, varying a subset of parameters at each
step, and and a Gaussian CTPDF, $q(\vec{\theta}'|\vec{\theta},
\mathcal{M})$ centered on $\vec{\theta}$ with a covariance matrix
$I\vec{\beta}$, where $I$ is the identity matrix and $\vec{\beta}$ is
a vector of scale parameters, $\beta_\nu$.  Throughout this paper, we
use the index $\nu$ to distinguish the model parameters.  The Gibbs
sampler specifies that only a subset of $\vec{\theta}$ is altered at
each step of the Markov chain.  While there are several variations, we
choose which parameters are to be altered in the next trial state
according to randomly generated permutations of the model parameters.
At each step one function of the model parameters
($u_\mu(\vec{\theta})$) is chosen to be updated using,
\begin{equation}
q(u_\mu(\vec{\theta}') | u_\mu(\vec{\theta}), \mathcal{M} ) = \frac{1}{\sqrt{2\pi \beta_\mu^2}} \exp \left[ -\frac{\left[u_\mu(\vec{\theta}')-u_\mu(\vec{\theta})\right]^2}{2\beta_\mu^2} \right]
\label{eqnCandTransProb}
\end{equation}
for valid $\vec{\theta}'$ (i.e., if the model dictates that
$\vec{\theta}'_\nu$ be positive definite, then trial states with
negative $\vec{\theta}'_{\nu}$ are rejected).  We use the index $\mu$
to distinguish the different types of steps, and the index $i$ to
indicate the $n$-th step of the Markov chain.  Each $\beta_\mu$ is a
parameter which controls the scale for the steps based on the quantity
indicated by $\mu$.  In Paper I, we used $\vec{u}(\vec{\theta}) =
\vec{\theta}$, i.e., we only took steps in the model parameters.  In
\S 4 of this paper, we will present several alternative CTPDFs that
can significantly improve the computational efficiency of the MCMC
algorithm by reducing the number of steps required before a Markov
chain can be used for inference.

\subsection{Summary of Previous Results}

The techniques of Bayesian inference and Markov chain Monte Carlo have
previously been applied to analyzing the radial velocity observations
of several extrasolar planetary systems.  Originally, Paper I
presented a complete algorithm for calculating the posterior
probability distributions for orbital parameters based on radial
velocity observations.  Paper I demonstrated several shortcomings of
the conventional estimates of parameter uncertainties based on
bootstrap-type resampling.  It found that Bayesian analyses, were
particularly important for planets where the orbital period is
comparable to the duration of observations.  Driscol (2006) is
performing a systematic study to compare uncertainty estimates made
with bootstrap-style resampling and uncertainty estimates made with
MCMC. Gregory (2005a) used Bayesian inference and MCMC to reanalyze
the observations of one particular system (HD 73526), and found two
alternative orbital solutions, one of which had a larger posterior
probability than the originally published orbital solution.  This
demonstrated another shortcoming of frequentist methods that are based
on only the maximum likelihood solution, rather than the posterior
probability distribution.  Ford, Lystad \& Rasio (2005) reanalyzed the
observations of the three planets orbiting $\upsilon$ Andromedae to
derive improved constraints on the orbital parameters and perform a
dynamical analysis (Ford et al.\ 2005).  Gregory (2005b) used an
improved algorithm to analyze another system (HD 208487) and claim
that the present data show a $\sim95\%$ probability for a second
planet.

\subsection{Motivation for this work}

Each of the above studies using MCMC has demonstrated advantages of
Bayesian inference for analyzing radial velocity observations of
extrasolar planets.  However, significant obstacles remain.  For
several of the systems analyzed in Paper I, the model parameters were
highly correlated and caused the Markov chains to converge slowly.
Convergence was particularly problematic for multiple planet systems,
so Paper I included an analysis of only one multiple planet system (GJ
876).  While we had attempted to include other multiple planet
systems, slow convergence rates and computational limitations
prevented us from being sufficiently confident that the other Markov
chains had converged to include them in the final paper.  Gregory
(2005a) took the more risky approach of basing inferences on Markov
chains which showed obvious signs of non-convergence, but argued that
the main conclusions were robust.  These studies illustrate the
importance of improving the computational efficiency of MCMC
algorithms for analyzing radial velocity data.

Since Paper I, we have dramatically improved the computational
efficiency of our MCMC algorithm by introducing alternative CTPDFs.
Our improved CTPDFs allow for the very rapid analysis of the typical
single planet systems, as well as the practical application of
Bayesian inference and MCMC to multiple planet systems.  For example,
incorporating just a few of these optimizations permitted the
dynamical analysis of the $\upsilon$ Andromedae system (Ford et al.\
2005).

\subsection{Outline}

In this paper we describe several refinements to the algorithms
presented in Paper I.  
In \S 2 we present our physical model of the planetary system and
observations.  We include modifications to the model that allow the
MCMC technique to be applied to systems with a significant amount of
stellar ``jitter'' and/or additional unknown planetary companions.  We
also present a method for testing the sensitivity of posterior
distribution functions to the assumption that the observational
uncertainties are normally distributed.
In \S 3 we provide precise descriptions of a few important technical
issues, including choosing priors, choosing step sizes, testing
for non-convergence, and deciding when to stop calculating Markov chains.
In \S4 we describe modifications to the original MCMC algorithm
presented in Paper I.  In particular, we present several alternative
choices for the CTPDF, $q(\vec{\theta}'|\vec{\theta})$, and
demonstrate the improved efficiency by applying the new sampling
algorithms to several different types of simulated planetary systems.
Of particular interest, we make practical suggestions for the choices
of CTPDFs in \S\ref{SRec}.
In \S 5, we discuss the application to multiple planet systems.
In \S 6, we demonstrate our improved algorithms on a few examples of
actual planetary systems.
In \S7, we summarize our conclusions and discuss areas for future
research.

\section{Model}

Here we present our model for the planetary system and the observations,
including several sources of noise.

\subsection{Model of Planetary System}

Since we are interested in planetary mass bodies, the dynamics is
described by the gravitational interactions of several point mass
bodies.  Specifying the mass and six phase space coordinates of each
body at a specified time would provide a complete description of the
system.  Shifting into the center of mass frame can eliminate one set
of phase space coordinates.
In practice, it is convenient to choose the osculating Keplerian
orbital elements (orbital period, $P$, orbital eccentricity, $e$,
inclination relative to the plane of the sky, $i$, argument of
periastron measured from the plane of the sky, $\omega$, longitude of
ascending node, $\Omega$, and mean anomaly, $M$) in Jacobi
coordinates.  The observed stellar velocity is the sum of the line of
sight velocity of the center-of-mass and the projection of the reflex
velocity due to any planetary companions onto the line of sight.  For
multiple planet systems, it can be important to use complete n-body
simulations to model the planetary motions accurately (e.g., GJ876;
Laughlin et al.\ 2005; Rivera et al.\ 2005).  However, in many cases,
the mutual planetary perturbations are negligible on time scales
comparable to the duration of observations.  In such cases, the radial
velocity perturbations due to a multiple planet system can be modeled
as the linear superposition of multiple unperturbed Keplerian orbits.

For a planet on an unperturbed Keplerian orbit, the mean anomaly is
the only one of the Keplerian orbital elements that changes with time.
The perturbation to the stellar radial velocity ($\Delta v_{*,p}$) due
to a planet on a Keplerian orbit is given by
\begin{equation}
\label{rveqn}
\Delta v_{*,p}(t) = K_p \left[ \cos\left(\omega_p+T_p\right) + e_p \cos(\omega_p) \right] 
\end{equation}
where $p$ labels the planet, $K$ is the velocity semi-amplitude and
$T$ is the true anomaly, which implicitly depends on time.  The true
anomaly ($T$) is related to the eccentric anomaly ($E$) via the
relation
\begin{equation}
\label{true anomaly}
\tan \left( \frac{T}{2} \right) = \sqrt{\frac{1+e}{1-e}} \tan \left( \frac{E}{2} \right).
\label{EccTrueEqn}
\end{equation}
The eccentric anomaly is related to the mean anomaly ($M$) via Kepler's equation
\begin{equation}
\label{KeplerEqn}
E(t) - e \sin\left(E(t)\right) = M(t) - M_o = \frac{2\pi}{P} \left( t - \tau \right)
\end{equation}
where $M_o$ is a constant, the orbital phase at $t=\tau$.  
The velocity semi-amplitude can be related to the planet mass, $m$,  by
\be
K = \frac{m \sin i}{M_*} \left( \frac{2 \pi G M_*}{P} \right)^{1/3} \left(1-e^2\right)^{-1/2} \left(1+\frac{m}{M_*}\right)^{-2/3} 
\label{EqnK}
\ee
where $M_*$ is the stellar mass and $G$ is the gravitational constant.

Unfortunately, radial velocity observations alone do not measure the
orbital inclination relative to the plane of the sky ($i$), the
longitude of ascending node ($\Omega$) or the planet mass.  The
minimum mass ratio, $m_{\min}/M_* = m \sin i/M_*$ can be determined
from Eqn.\ \ref{EqnK} iteratively.  When only radial velocity
observations are used to constrain the orbital elements, we simply
exclude $i$ and $\Omega$ from the set of model parameters for planet
$p$, $\vec{\vartheta}_p = ( \log P_p, \log K_p, e_p, \omega_p, M_{o,p}
)$.  However, when additional constraints are available (e.g.,
astrometric or dynamical), these can be constrained and should be
included in $\vec{\vartheta}_p = ( \log P_p, \log K_p, e_p, \omega_p,
M_{o,p}, \cos i_p, \Omega_p )$.

\subsection{Model of Observations}

In radial velocity surveys, the velocity of the central star is
precisely monitored for periodic variations which could be caused by
orbiting companions.  Each individual observation can be reduced to an
estimate of the observational uncertainty ($\sigma_{k,\mathrm{obs}}$)
and a measurement of the star's radial velocity
($v_{*,\mathrm{obs}}(t_k,j_k)$), where $t_k$ is the time of the $k$th
observation and $j_k$ specifies the observatory and spectrometer used
for the $k$th observation.
Because each radial velocity measurement is based on calculating the
centroid of thousands of spectra lines and averaged over hundreds of
sections of the spectrum, the observational uncertainties of most
current echelle based radial velocity surveys can be accurately
estimated and are nearly Gaussian (Butler et al.\ 1996).
If the observational data ($\vec{d}$) were generated by the model
specified by $\vec{\theta}$, then the probability of drawing the
observed values is
\begin{equation}
p(\vec{d}|\vec{\theta},  \mathcal{M}) = \Pi_{k} \frac{1}{\sqrt{2\pi}\sigma_k} \exp \left[ -\frac{\left(d_{k,\theta}-d_k\right)^2}{2\sigma_k^2} \right],
\label{eqnpdx}
\end{equation}
assuming that the errors in individual observations are normally
distributed and uncorrelated.  Here $d_k \equiv v_{*,\rm
obs}(t_k,j_k)$ is the observed velocity at time $t_k$ as measured by
observatory and spectrometer $j_k$, $d_{k,\theta} =
v_{*,\vec{\theta}}(t_k,j_k)$ is predicted velocity according to the
model parameters $\vec{\theta}$, and $\sigma_k$ is the uncertainty in
the $k$th observation.
If there are no sources of noise other than the measurement
uncertainties, $\sigma_{k,\rm obs}$, then we can use Eqn.\
\ref{eqnpdx}, setting $\sigma_k = \sigma_{k,\rm obs}$.
Since the observational uncertainties are all assumed to be Gaussian
and uncorrelated, it is convenient to use the $\chi^2$ statistic,
\begin{equation}
\label{chisqeqn}
\chi^2 = \sum_k \left[ v_{*,\vec{\theta}}\left(t_k,j_k\right) - v_{*,\rm obs}\left(t_k,j_k\right] \right)^2 / \sigma_{k}^2.
\end{equation}
If the planets are on non-interacting Keplerian orbits, then 
\be
v_{*,\vec{\theta}}(t,j) = C_j + \sum_p \Delta v_{*,p}(t),
\ee
where $C_j$ is the velocity constant for the observatory/spectrograph
labeled by subscript $j$, and $\Delta v_{*,p}(t)$ is given by
Eqn. \ref{rveqn}.  While there is a single mean line of sight velocity
of the center of motion, it is important to use separate constants,
$C_j$ for each observatory/spectrograph pair due to potential
differences in the calibrations.  In the next section we discuss
observational uncertainties and present reasons why the values of
$\sigma_k$ may depend on the model parameters.

\subsection{Observational Uncertainties and Other Sources of Noise}

In Paper I, we analyzed chromospherically inactive stars and assumed
that the uncertainties in individual observations were well estimated.
Therefore, we set $\sigma_k = \sigma_{k,\rm obs}$. However, for many
stars the distribution of $\chi^2$ calculated with the actual data and
models drawn from the posterior distribution was larger than would be
expected if the model were correct.  Even if the observational
uncertainties are perfectly estimated, there could be additional
causes of radial velocity variations that are not included in the
model.  We explore the two most important causes of ``jitter'' below.
Then, we modify the methods described above to allow for an additional
source of observational uncertainty of unknown magnitude.

\subsubsection{Stellar Jitter}

For many stars the small chromospheric emission suggests that the
observational uncertainties are dominated by photon noise.  Indeed,
the quoted uncertainties in the radial velocity observations have been
demonstrated to be quite accurate for some chromosphically inactive
stars with no signs of planets (Wright 2005).  However, for other
stars, the radial velocity observations can be contaminated by stellar
``jitter'' due to stellar phenomena such as stellar oscillations,
convective motions, starspots.

The true distribution for the stellar jitter is unknown.  One of the
primary sources of stellar jitter is believed to be stellar p-mode
oscillations.  Indeed, such oscillations have been detected for a few
stars by taking a rapid series of high precision radial velocity
observations (e.g., Kjeldsen et al.\ 2003; Butler et al.\ 2004).  We
treat the jitter as an additional source of Gaussian noise with
variance $\sigma_{\rm jitter}^2$.  This is an excellent approximation
for two reasons.  First, the stellar jitter will be convolved with the
Gaussian observational uncertainty.  Second, stellar oscillations
would result in a periodic signal of finite amplitude, so the jitter
distribution would be symmetric and have smaller tails than a Gaussian
distribution with the same variance.  Thus, we expect that
approximating the jitter as an additional source of Gaussian noise
will be highly accurate.  To test this assumption, we have analyzed a
large number of observations of the star $\tau$ Ceti taken by the
California \& Carnegie planet search (Fischer, private communication).
We find that the distribution of the velocities for $\tau$ Ceti is
consistent with a normal distribution, and conclude that the above
treatment of stellar jitter is adequate for the purposes of this
paper.

\subsubsection{Additional Unseen Companions}

Another potential source for excess scatter in the residuals is the
presence of additional undetected companions.  When the number of
observations is small ($\lo50$), it can be possible to detect a
significant increase in the scatter, even if it is not possible to
identify an orbital solution which significantly reduces $\chi^2$
(Cumming et al.\ 2003).  Since the amplitude, eccentricity, and
argument of pericenter of any undetected companions is unknown, the
exact distribution of these perturbations is unknown.  While it is
theoretically possible to calculate the distribution of the
perturbations based on the prior distributions for $\log K$, $e$, and
$\omega$, we note that this distribution is symmetric and has tails
that are less than that of a normal distribution with the same
variance.  Additionally, the distribution of the perturbations will be
convolved with the Gaussian observational uncertainties.  Therefore,
the distribution of the perturbations can be conveniently approximated
as normal with zero mean and variance $\sigma_{\rm unseen}^2$.  We can
easily incorporate the perturbations due to both stellar ``jitter''
and unseen companions by using Eqn.\ \ref{eqnpdx} and setting
$\sigma_k^2 = \sigma_{k,\rm obs}^2 + \sigma_+^2$, where $\sigma_+^2 =
\sigma_{\rm jitter}^2 + \sigma_{\rm unseen}^2$.

\subsubsection{Correlated Residuals}

The above treatment of stellar jitter and undetected companions
assumes that the velocity perturbations due to these effects are
uncorrelated in time.
If the stellar jitter is due to convection or stellar oscillations,
then we expect significant correlation over minute time scales, but
not days.  On the other hand, if star spots introduce significant
stellar jitter, then there could be correlations on time scales
comparable to the rotation period (ones to tens of days).
If an undetected planet has an orbital period of just a few days, then
the residuals for observations on consecutive nights could have a
significant correlation, but observations made during consecutive
months would typically have a negligible correlation.  In this case
the posterior distribution for the planets detected with significantly
longer periods will generally be near the actual parameters but with
an increased dispersion due to the ``noise'' from the undetected
planet.  However, if an undetected companion has an orbital period
longer than the typical spacing of observations, then the residuals at
consecutive observation times will be correlated, possibly resulting
in a systematic effect on the posterior distribution for the orbital
parameters of the detected planets.  The magnitude of the velocity
perturbations should be small (since the planet is presumed to be
undetected), so in many cases the effect on the orbital parameters for
the detected planets is still expected to be small.  However, in some
cases the perturbations of an undetected planet can be largely
compensated for by a change in the orbital parameters of the detected
planets (e.g., the inner planet in a 2:1 mean motion resonance can
have an effect similar to an eccentricity for the outer planet).  In
such cases, the resulting posterior distribution may be biased by the
unseen companion.  Simply increasing the effective observational
uncertainty does not account for the correlation between observations
due to the unknown companion.  This potential bias could be removed by
including the possibility of additional unknown companions in the
model and marginalizing over their orbital parameters.  Unfortunately,
this would be computationally quite demanding, since the allowed
orbital parameters for the unknown companion will likely span the
entire allowed parameter space.  Nevertheless, if the radial velocity
signature of a distant companion is even marginally detected, then the
distant companion should be properly incorporated into the model, even
if the observations provide only limited constraints on the orbital
parameters.  This provides yet another motivation to improve the
efficiency of MCMC algorithms for analyzing radial velocity data.

\subsection{Models with Noise of Unknown Magnitude}

The radial velocity perturbations due to stellar jitter and undetected
planets can be incorporated into our models by assuming that the
jitter causes a perturbation to the radial velocity at each
observation.  If we assume that the perturbations due to ``jitter''
are uncorrelated and drawn from a normal distribution with zero mean
and variance $\sigma_+^2$, then the expectation of
$v_{*,\vec{\theta}}(t_k,j_k)$ is unchanged, but the distribution of
the residuals, $v_{*,\rm obs}(t_k,j_k) - v_{*,\vec{\theta}}(t_k,j_k)$,
is normal with a variance equal to $\sigma_{k,\rm obs}^2+\sigma_+^2$.
The probability of the observations ($\vec{d}$) given the model
($\vec{\theta}$) is given by Eqn.\ \ref{eqnpdx} by setting $\sigma_k^2
= \sigma_{k,\rm obs}^2 + \sigma_+^2$ (Gregory 2005a).  If the value of
each $\sigma_k$ were known {\em a priori}, the terms outside the
exponential are constant from model to model, they cancel when
considering the ratio $p(\vec{d}|\vec{\theta}',
\mathcal{M})/p(\vec{d}|\vec{\theta}, \mathcal{M})$ for calculating the
acceptance probability with in Eqn.\ \ref{AlphaEqn}.  When the value
of $\sigma_+$ is not known {\em a priori}, it must be estimated along
with the other parameters in $\vec{\theta}$.  The terms outside the
exponential of Eqn.\ \ref{eqnpdx} are no longer the same for all
models.  Therefore, when a candidate transition step varies
$\sigma_+$, the acceptance rate will depend on the values of
$\sigma_+$ for the two models, both due to the effect on
$\chi^2(\vec{\theta}')-\chi^2(\vec{\theta})$ and due to the ratio
\be
\frac{\Pi_k \left( \sigma^2_{k, \rm obs} + \sigma_+^2 \right)^{1/2}
}{\Pi_k \left( \sigma^2_{k, \rm obs} + {\sigma'}_+^2 \right)^{1/2}}.
\ee
This latter factor penalizes models for which $\sigma_+$ would result
in a larger dispersion of velocity residuals than are contained in the
actual data.  For observational data which requires a significant
noise component, a small value of $\sigma_+$ is penalized by the
effect on $\chi^2$.

Since $\sigma_+$ is a scale parameter, it would be natural to use the
non-informative prior $p(\sigma_+) \sim \sigma_+^{-2}$.  However, a
prior for $\sigma_+$ that is flat on $\log \sigma_+$ is improper,
i.e., the prior probability distribution function is not normalizable.
If a significant value of $\sigma_+$ is required to match the
observational data, then the posterior distribution will be
normalizable.  If the observational data requires that any noise
source be small compared to the observational uncertainties,
$\sigma_{k, \rm obs}$, then the posterior distribution if likely to be
non-normalizable, but the value of the parameter $\sigma_+$ will not
make a significant difference in the model predictions.  In such a
case, new Markov chains could be calculated with $\sigma_+$ held fixed
at $\sigma_+=0$ to provide a properly normalized posterior
distribution.  We can avoid this complication by using a normalized
prior for $\sigma_+$.  For example, we could impose sharp lower and
upper limits, $-5 \le \log \sigma_+ \le 7$.  An alternative is to use
the Jefferies prior,
\be
p(\sigma_+) = \left(\sigma_{+,o}+\sigma\right)^{-1}\left[\ln\left(\frac{\sigma_{+,o}}{\left(\sigma_{+,o}+\sigma\right)}\right)\right]^{-1},
\ee
which is also properly normalized (Gregory 2005a).

Finally, we note that allowing the model to include a noise source
with unknown magnitude can have practical benefits when a Markov chain
is started far from the best-fit solutions.  If $\chi^2$ is calculated
assuming only observational uncertainties, then the $\chi^2$ surface
is often very ``rough'' with many local minima and maxima scattered
throughout parameter space.  However, when an unknown noise source is
included in the model and the current state has a large $\chi^2$, then
the Markov chain will increase $\sigma_+$ so that $\chi^2$ is
comparable to the number of degrees of freedom (number of observations
minus number of free parameters).  This will result in smoothing out
the $\chi^2$ surface (at constant $\sigma_+$) and allow the Markov
chain to explore the parameter space more quickly.  The Markov chain
will begin to decrease the values of $\sigma_+$ as the Markov chain
settles in near the best-fit parameters.  This behavior is similar to
simulated annealing, but does not require choosing a cooling schedule
in advance.  Thus, including a noise source of unknown magnitude can
have the additional benefit of reducing the number of steps required
for a Markov chain to go from an initial state to the region of
parameter space near the best-fit solution.  This is particularly
useful when the Markov chains will be tested for non-convergence
using tests which require that the Markov chains be started from
widely dispersed initial conditions, such as the Gelman-Rubin test
(see \S3.3).

\subsubsection{Non-Gaussianities \& Robust Statistics}

In the above treatment we have assumed that the distribution for the
unknown radial velocity perturbations is normal.  Since the actual
distribution of ``stellar jitter'' (or other poorly understood noise
sources) is unknown, one might worry about the robustness of
conclusions based on the assumption of normality.  Here we introduce a
method for testing the sensitivity of conclusions to this assumption.
We replace the normal distribution, $N(0,\sigma_k)$ for each of the
velocity residuals with a Student's t distribution with $\upsilon$
degrees of freedom ($t_\upsilon(0,\sigma_k)$).  In the limit that
$\upsilon \rightarrow \infty$, the distribution
$t_\upsilon(0,\sigma_k)$ approaches the normal distribution,
$N(0,\sigma_k)$.  For $\upsilon\le2$ the tails of the $t_\upsilon$
distribution are so large that the variance is undefined.

To implement this alternative assumption about the distribution of
errors, we replace Eqn.\ \ref{eqnpdx} with
\begin{equation}
p_{\upsilon}(\vec{d}|\vec{\theta}) = \prod_{k} \frac{\Gamma((\upsilon+1)/2)}{\Gamma(\upsilon/2)\sqrt{\upsilon\pi}\sigma_k} \left[ 1+ \frac{1}{\upsilon}\left(\frac{v_{*,\vec{\theta}}-v_{*, \rm obs}}{\sigma_k}\right)^2 \right]^{-(\upsilon+1)/2},
\label{EqnTDist}
\end{equation}
where both $v_{*,\vec{\theta}}$ and $v_{*,obs}$ are functions of $t_k$
and $j_k$.  It is then possible to compute multiple posterior
distributions parameterized by $\upsilon$.  In principle, $\upsilon$
could be treated as an unknown hierarchical model parameter and
directly estimated from the observational data with the standard
Bayesian techniques.  In most cases, the posterior probability
(marginalized over all model parameters other than $\upsilon$) 
reveals a lower bound on $\upsilon$, $\upsilon_{\min}>2$, since
$\lim_{\upsilon\rightarrow2} \var(t_\upsilon(0,\sigma_k)) = \infty$.
In practice, a very large number of observations is required to
characterize the wings of the probability distribution for the
observational data (Gelman et al.\ 2003).  Therefore, in typical
cases, it is not possible to place an upper bound on $\upsilon$,
since $\lim_{\upsilon\rightarrow\infty} t_\upsilon(0,\sigma_k) \sim
N(0,\sigma_k)$, corresponding to the case where observational
uncertainties are due to photon noise alone.  Unless there
is a good reason for believing that the velocity residuals follow a
$t$-distribution with a particular value of $\upsilon$, the posterior
distributions calculated using a $t_\upsilon$ distribution
($p_\upsilon(\vec{\theta}|\vec{d})$), should not be regarded as a
robust estimate of the posterior distribution.  Rather, we recommend
that researchers analyze posterior distributions with multiple values
of $\upsilon$ to determine the sensitivity of inferences made to the
assumed distribution for the velocity residuals (e.g., as in Ford et
al. 2005).  If there are significant differences between inferences
based on $p_{\infty}(\vec{\theta}|\vec{d})$ and
$p_{\upsilon_{\min}}(\vec{\theta}|\vec{d})$, then one should realize
that the inference depends on the assumed distribution for the
velocity residuals.  If the same inferences could be made from either
$p_{\infty}(\vec{\theta}|\vec{d})$ or
$p_{\upsilon_{\min}}(\vec{\theta}|\vec{d})$, then the inferences can
be considered robust to the assumptions that observational
uncertainties and other sources of noise are normally distributed.

\section{Technical Issues Relating to MCMC}

We have summarized the essential elements of the MCMC algorithm in
\S1.2 and our models for analyzing radial velocity observations of
extrasolar planetary systems in \S2.  Here we discuss a few technical
issues related to the practical implementation of our MCMC algorithm.

\subsection{Choice of Priors}

We choose a uniform prior in each of the parameters in $\vec{\theta} =
( \vec{\vartheta}, \vec{C}, \log\sigma_+ )$.  Remember that
$\vec{\vartheta}_p = ( \log P_p, \log K_p, e_p, \omega_p, M_{o,p} )$
if radial velocity observations are the sole constraint or
$\vec{\vartheta}_p = ( \log P_p, \log K_p, e_p, \omega_p, M_{o,p},
\cos i_p, \Omega_p )$ if additional constraints are available, where
$P$ is the orbital period, $K$ is the velocity semi-amplitude, $e$ is
the orbital eccentricity, $\omega$ is the argument of periastron, and
$M_o$ is the mean anomaly at the chosen epoch, $\tau$.  The constant
velocity offsets, $C_j$, are a location parameter, so the natural
choice for an uninformative prior is uniform in $C_j$.  We have
already discussed the choice of priors for $\sigma_+$ in \S2.4.
Formally, these priors for $\log P_p$, $\log K_p$, are $C_j$ are
improper, i.e., non-normalizable.  We solve this by imposing sharp
limits on these parameters, $\log P_{\min} \le \log P \le \log
P_{\max}$, $\log K \le \log K_{\max}$, and $C_{\min} \le C \le
C_{\max}$.  The choice of limits can be based on other physical
constraints (e.g., ability of planet to survive at very short/long
orbital periods, a sufficiently massive companion could be detected in
the spectra of the primary, and the escape velocity from the galaxy).
However, in most cases, these limits will be chosen sufficiently far
away from the viable solutions that their exact choice will not affect
the shape of the posterior distribution,
$p(\vec{\theta}|\vec{d},\mathcal{M})$.  While the possibility of
arbitrarily small velocity amplitudes prevents a physical
justification for a $K_{\min}$, clearly it is not possible to detect
or constrain the orbital parameters of a planet with $K$ much less
than the $\sigma_{\rm{ave}} = 1/\sum_k \sigma_k^{-2}$ without many
observations.  We suggest $K_{\min} = K_{0.5}/2$, where $K_{0.5}=
\sigma_{\rm{ave}} \sqrt{50/(N_{obs}-N_{param})}$ approximates the
amplitude for which there is a 50\% probability of detecting a planet
based on the simulations of Cumming (1999).  For systems where a
planet is clearly detected, the posterior will not be sensitive to our
assumptions about the prior for small values of $K$, but for planets
which are marginally detected, this choice may become significant.  A
reasonable choice of a normalizable prior is a Jeffery's prior $p(K) =
(K+K_o)^{-1} \left[\log(1+K_{\max}/K_o)\right]^{-1}$, where $K_o
\simeq K_{0.5}$ (Gregory 2005a).  However, if the posterior
distribution includes values of $K\simeq K_o$, then one should be
particularly careful to check how sensitive any conclusions are to the
choice of $K_o$.  If we were to use flat priors with strict limits
($p(\vec{\theta}| \mathcal{M})\sim 1$), then the posterior
distribution, $p(\vec{\theta}|\vec{d}, \mathcal{M})$, is proportional
to $p(\vec{d}|\vec{\theta}, \mathcal{M})$.  For the remainder of this
paper, we use Jefferies priors for $K$ and $\sigma_+$, so
\be
p(\vec{\theta}|\vec{d}, \mathcal{M}) \sim 
\frac{p(\vec{d}|\vec{\theta}, \mathcal{M})}{\sigma_{+,o} + \sigma_+}
\prod_p \left(K_o+K_{p}\right)^{-1},
\ee
provided that $\log P_{\min} \le \log P_p \le \log P_{\max}$, $\log
K_p \le \log K_{\max}$, $0 \le e_p \le 1$, $0 \le \omega_p \le 2\pi$,
$0 \le M_{o,p} \le 2\pi$, $C_{\min} \le C_j \le C_{\max}$, and
$\sigma_+ \le \sigma_{+,\max}$.  In practice, the reported stellar
velocities are typically measured relative to the velocity of the star
at the time of a template observation (e.g., a high-resolution
observation made without an iodine cell), so the magnitude of each
$C_j$ should be no more than the amplitude of the radial velocity
variations.  Since comparable mass stellar binaries would be easily
detected by the presence of a second set of spectral lines, $C_{\max}
= - C_{\min} = 100$ km/s is certain to include all allowed orbital
solutions, including the possibility spectroscopic binaries.  In
practice, much smaller values of $C_{\max}$ could be adopted (e.g.,
based on the range of the observed radial velocities).  While this
might be useful for some analysis algorithms, it is not necessary for
the techniques used in this paper.

\subsubsection{Comparison to Paper I}

In paper I, we used $\vec{\vartheta} = \left( \log P, \log K, e \sin
\omega, e \cos \omega, M_o\right)$ for calculating our Markov chains, 
so the Markov chains sampled from a posterior distribution which
assumed 
\bea
& p(e\sin \omega, e\cos \omega) & d(e\sin\omega) d(e\cos\omega)  \nonumber \\
&& =  1/\pi d(e\sin\omega) d(e\cos\omega)  \nonumber \\
&& =  p(e,\omega) de d\omega = e/\pi  de d\omega 
\label{EqnHkToEomega}
\eea
We then used importance sampling to compute posterior distributions
which corresponded to the desired priors, $p(e,\omega)=1/2\pi$ (Gilks
et al.\ 1996).  We made this choice in paper I, since taking steps in
$e\sin\omega$ and $e\cos\omega$ was more efficient than taking steps
in $e$ and $\omega$ for low eccentricity systems.  However, this
choice also decreased the computational efficiency for long period
systems where long period and high eccentricity solutions could not
yet be fully excluded.  In this paper, we will introduce several
additional CTPDFs that allow our Markov chains to efficiently jump
from one state to another for both low and high eccentricity models,
using a single choice of prior.  This eliminates the need for
importance sampling to correct for originally calculating a Markov
chain using a different prior than is desired for making inferences.

\subsubsection{Alternative Priors}

From the theorist's perspective, it might be desirable to assume a
prior which is uniform in $\log \mu_{\min} = \log (m_{\min}/M_*)$,
where $m_{\min}$ is the minimum allowed mass for given values of $P$,
$K$, $e$, and $M_*$ and can determined by setting $\sin i =1$ in Eqn.\
\ref{EqnK}.  However, radial velocity perturbations have a more direct
dependence on $K$, the velocity semi-amplitude than the planetary
mass, $m$.  For this reason we have constructed Markov chains using a
prior which is based on $K$.

If it were important to assume a prior based on $\mu_{\min}$, then
importance sampling could be applied to an existing Markov chain.  The
Jacobian of the transformation is $J =
(1+\mu_{\min})/(1+\mu_{\min}/3)$.  Thus, we could weight each state of
a Markov chain assuming a prior uniform on $\log P$, $\log K$, and $e$
by $J$ to obtain an estimate of the posterior probability distribution
based on a prior which is uniform on $\log P$, $\log \mu_{\min}$ and
$e$.  If we desire a sample from the posterior using this alternative
prior, then we can apply rejection sampling to the states calculated
in the Markov chain based on $\log K$.  In practice, $\mu_{\min}\ll 1$
for planetary mass companions, so $J \simeq 1+ 2\mu_{\min}/3 +
O(\mu_{\min}^2) \simeq 1$ and there is very little difference.  In
fact, if the posterior distribution is sensitive to whether the prior
is in terms of $K$ or $\mu_{\min}$, then either the companion is of
nearly stellar mass or this is an indication that the data are
inadequate to constrain either $K$ or $\mu_{\min}$.

A similar method can be used to sample from a prior that is uniform
on $\log a_{\min}$ (the semi-major axis if $\sin i =1$) rather than
$\log P$.  This transformation has a Jacobian of $J = \left[2 \left( 6
+ 3 \mu_{\min}\right)\right]\left[3 \left(6+2\mu_{\min}\right)\right]
\simeq 1 + \mu_{\min}/6 +O(\mu_{\min}^2)$.  Again, the two priors are
very similar for planetary mass companions and significant differences
in the posteriors are either a stellar mass companion or an indication
that the data are inadequate to constrain either $\log P$ or $\log a$.
While a reasonable argument could be made in favor of either of these
priors, we choose to continue using a prior based on $\log P$ for
planetary mass companions, as the differences are negligible.

\subsection{Adaptive Step Size Algorithm}

In Paper I, we computed Markov chains using a wide range of step
sizes.  We dismissed the possibility of non-convergence by verifying
that Markov chains constructed with these widely varying step sizes
all resulted in similar posterior distributions.  Unfortunately, the
rate of convergence depends sensitively on the step sizes chosen, so
computing Markov chains with a wide range of step sizes necessarily
results in some chains being very inefficient.  In this paper, we
suggest abandoning the practice of construct multiple chains with
widely varying step sizes, in favor of running more chains with an
appropriately chosen step size.  Theoretical results (which assume
$p(\vec{\theta} | \vec{d}, \mathcal{M})$ is normally distributed)
suggest that the best choice for each $\beta_{\mu}$ would result in an
acceptance rate of $\simeq0.44$ (Gelman et al.\ 2003).  This led us to
develop an algorithm to automatically optimize the choice of step
sizes, $\vec{\beta}$.  While the algorithm outlined below has not been
carefully optimized, we find that it provides acceptable results.

At first, we guess reasonable values for the step scale parameters,
$\vec{\beta}$.  We then begin running a Markov chain and monitor the
acceptance rates, $\vec{\psi}$, for each type of step separately.  We
periodically compare each component of $\vec{\psi}$ to the desired
acceptance rate, $\psi_o =0.44$, to determine if the corresponding
elements of $\vec{\beta}$ should be altered.  We treat update the
$\beta_\mu$ when $(\psi_\mu-\psi_o)^2 > s_\mu^2 \psi_o (1-\psi_o) /
N_\mu$, where $s_\mu^2$ is a parameter that determines how frequently
$\beta_\mu$ should be updated ($s_\mu^2$ is initially set to 2), and
$N_\mu$ is the total number of steps taken using that type of step
since the last update of $\beta_\mu$ (including repeats of the same
state when a proposed step is rejected).  When we determine that
$\beta_\mu$ is to be updated, we multiply $\beta_\mu$ by
$(\psi_\mu/\psi_o)^\phi$, where $\phi=1$ for $0.5 \psi_o < \psi_\mu$,
$\phi=1.5$ for $0.2 \psi_o < \psi_\mu \le 0.5 \psi_o$, and $\phi=2$
for $0.1 \psi_o < \psi_\mu \le 0.2 \psi_o$, but we never reduce
$\beta_\mu$ by more than a factor of 100.  We impose an upper limit of
$\beta_\mu = 4\pi$ for angular variables.  After updating an element
of $\vec{\beta}$ we resume computing a Markov chain from the final
state of the Markov chain prior to updating $\vec{\beta}$.
Technically, changing any one of the step sizes destroys the Markov
property.  Therefore, we must not use the previous states in the chain
for inference.  Nevertheless, it is more efficient to retain the
counts of accepted and rejected steps for each step type (that did not
have its scale parameter updated).  We find that this produces
acceptable results, for the purposes of choosing the step sizes.  This
process is typically repeated several times to determine an efficient
choice for each of the scale parameters in $\vec{\beta}$.  Any time
$\beta_\mu$ is increased and then consecutively decreased (or vice
versa), we increment $s_\mu^2$.  We stop adjusting the scale
parameters in $\vec{\beta}$ once each $\psi_\mu$ is within 10\% of
$\psi_o$ or the corresponding $\beta_\mu = 4\pi$ for an angular
variable.

\subsection{Tests for Non-convergence}

In Paper I we used the Gelman-Rubin statistics, $\widehat{R}$, to test
for non-convergence of Markov chains and to decide when the Markov
chains were suitable for making inferences about the orbital
parameters of a particular system.  The Gelman-Rubin statistic
compares the variance of any quantity, $z(\vec{\theta})$, estimated
from each of the individual Markov chains to the variance of the
estimates of the mean of $z(\vec{\theta})$ from the different chains.
Following Gelman et al.\ (2003), we consider the quantities $z_{ic}$
that can be calculated from the model parameters at each iteration
(indexed by $i$) of each Markov chain (indexed by $c$).  If we have
$N_c$ Markov chains each of length $L_c$, then we could estimate the
mean value of $z_{ic}$ based on each of the Markov chains as
\be
\bar{z}_{\cdot c} = \frac{1}{N_c} \sum_{i=1}^{N_c} z_{ic},
\ee
and the average of the variances of $z_{ic}$ within each chain as
\be
W(z) = \frac{1}{N_c}\sum_{c=1}^{N_c} \frac{1}{L_c-1} \sum_{i=1}^{L_c} \left( z_{ic}- \bar{z}_{\cdot c} \right)^2.
\ee
Similarly, we could estimate the mean value from the entire set of
Markov chains as
\be
\bar{z}_{\cdot\cdot} = \frac{1}{N_c} \sum_{c=1}^{N_c} \bar{z}_{\cdot c}, = \frac{1}{L_c N_c} \sum_{c=1}^{N_c} \sum_{i=1}^{L_c} z_{ic}
\ee
and the variance of the estimates of the single chain means by
\be 
B(z) = \frac{L_c}{N_c-1} \sum_{c=1}^{N_c} \left(\bar{z}_{\cdot c} - \bar{z}_{\cdot\cdot} \right)^2.  
\ee
Then we can estimate the variance of $z_{ic}$ by a weighted average of
$W(z)$ and $B(z)$,
\be
\widehat{\var^+}(z) = \frac{L_c-1}{L_c} W(z) + \frac{1}{L_c} B(z).
\ee
This is an unbiased estimator of $\var (z)$ either if the $N_c$
initial states of the Markov chains were selected from the target
distribution or in the limit $L_c\rightarrow\infty$.  If the $N_c$
Markov chains were started from a set of states which have a larger
variance than the posterior, then $\widehat{\var^+}(z)$ will initially
overestimate $\var(z)$, while $W(z)$ will underestimate it, since the
individual Markov chains may not yet have enough states for $z$ to
explore its whole range.  From these quantities, we form an estimate
of $\widehat{R}(z)$,
\be
\widehat{R}(z) = \sqrt{\frac{\widehat{\var^+}(z)}{W(z)}},
\ee
that is the factor by which the scale of the estimate of the
distribution $p(z(\vec{\theta} | \vec{d}, \mathcal{M})$ could be
reduced by continuing to calculate longer Markov chains (Gelman et
al. 2003).  As the individual Markov chains approach convergence,
$\widehat{R}(z)$ approaches 1 from above.  In paper I, we required
that $\widehat{R}_\nu\le 1.1$ for each $\nu \in ( \log P, \log K, e
\sin \omega, e \cos \omega, M_o, C )$ before using a chain for
inference.

\subsection{Stopping Criteria}
\label{StopCriteria}
In \S 4 of this paper, we will present several modifications to the
algorithm of Paper I and compare the rates of convergence of Markov
chains calculated with different algorithms.  To facilitate
quantitative comparisons of a large number of Markov chains, we have
adopted an algorithmic stopping criteria which we will use to estimate
a stopping time, $N_{Stop}$.

After determining an appropriate set of step sizes, we begin
calculating $N_c = 10$ Markov chains for each system to be analyzed.
We periodically pause to calculate the Gelman-Rubin statistics,
$\widehat{R}(z_\rho(\vec{\theta}))$, for each of several variables
indexed by $\rho$.  Since we desire to accurately estimate,
$N_{Stop}$, we initially check after every 100 steps in each variable,
but then increase the time between tests so as to maintain
approximately 10\% precision in the estimate of $N_{Stop}$.  We
require that $\widehat{R}(z_\rho) \le 1.01$ for each $\rho$.  We also
calculate an estimate of the effective number of independent draws,
$\widehat{T}(z)$,
\be
\widehat{T}(z) = L_c N_c \min \left[ \frac{\widehat{\var^+}(z)}{B(z)}, 1 \right].
\ee
This attempts to correct for the effects of autocorrelation within
each of the Markov chains on the total number of states in the Markov
chains (Gelman et al. 2003).  We require that
$\widehat{T}(z_\rho(\vec{\theta})) \ge 1000$ for each quantity that we
monitor (labeled by $\rho$).  When all the tests based on both
$\widehat{R}(z_\rho)$ and $\widehat{T}(z_\rho)$ are satisfied, we set
$N_{Stop}$ to be the number of steps in each Markov chain at this
time.

Since we repeatedly test our Markov chains with these criteria, there
is an increased chance that the values of $\widehat{R}$ and
$\widehat{T}$ will fluctuate below our thresholds by chance.
Therefore, once both criteria are met, we then continue computing the
Markov chains to increase the length of the Markov chains by
approximately 1\%, 2\%, 3\%, 4\% and 5\%.  and repeat all the tests at
each of these times.  If the criterion for any one of the
$\widehat{R}(z_\rho)$ or $\widehat{T}(z_\rho)$ are not met at any of
these times, then we discard the tentative value of $N_{Stop}$ and
continue computing each of the Markov chains.  Therefore, our final
stopping criteria requires $\widehat{R}(z_\rho)$ and
$\widehat{T}(z_\rho)$ satisfy the above criteria for all quantities
labeled by $\rho$ at all five consecutive tests, but we report the
number of steps before the first of the five consecutive set of passed
tests.

Special care is required for applying the Gelman-Rubin test to
variables which are angles.  If the mean values of a variable in two
chains are $\epsilon$ and $2\pi-\epsilon$, then the difference between
the mean values should be $2\epsilon$ not $2\pi-2\epsilon$.
Similarly, care must be taken when calculating the variance of an
angular variable within a Markov chain not to overestimate the
variance due to the particular representation of the angle.  To
circumvent this problem, for any angle, $\xi$, we use the mean and
variance of a ``standardized'' angle, $\xi_{\mathrm{std}}$.  First, we
define the notation $\left\{\xi\right\}$ to be a representation of the
angle $\xi$ that lies in the interval $[-\pi,\pi)$.  We consider the
values of $\left\{\xi-w\pi/8\right\}$, where $w$ ranges from 0 to 15,
and calculate their mean, $\overline{\left\{\xi-w\pi/8\right\}}$, and
variance, $\var(\left\{\xi-w\pi/8\right\})$.  We then set
$w^*$ to be value of $w$ which minimizes
$\var(\left\{\xi-w\pi/8\right\})$, and obtain an initial
estimate for the mean standardized angle, $\bar{\xi}_{\mathrm{std},1}
\equiv w^*\pi/8+\overline{\left\{\xi-w^*\pi/8\right\}}$.  We then
compute the mean and variance of
$\left\{\xi-\bar{\xi}_{\mathrm{std},1}\right\}$ to determine our final
estimates of the standardized mean
$\bar{\xi}_{\mathrm std} \equiv 
\bar{\xi}_{\mathrm{std},1} + \left\{ \xi - \bar{\xi}_{\mathrm{std},1} \right\} 
$
and the
standardized variance, 
$
\var(\xi_{\mathrm{std}}) \equiv 
\var\left(\xi- \bar{\xi}_{\mathrm{std},1} \right) 
$
 Note, that it is important to use a two pass method to calculate
these variances.

If we were constructing these Markov chains for the purpose of
inference we would not calculate the values of $\widehat{R}$ so
frequently or stop our Markov chains as soon as this criteria were
satisfied.  We do not claim that this set of tests is sufficient to
proof convergence.  Indeed, it is impossible to prove convergence and
only possible to prove non-convergence.  Therefore, the standard
practice is to perform multiple tests for non-convergence until one is
satisfied that the Markov chain is sufficiently close to convergence
to be used for inference (Gelman et al.\ 2003).  The above stopping
criteria is one example of such a suite of tests for non-convergence.
We believe that the above set of tests provides an adequate means of
quantitatively assessing the efficiency of the various CTPDFs which we
explore in \S 5.

\section{Improving the Efficiency of MCMC}

The computational efficiency of the basic MCMC algorithm presented in
Paper I was typically limited by high correlations between some
model parameters.  One obvious strategy for improving the efficiency of
MCMC would be to perform a principal components analysis (PCA) of the
states from a preliminary Markov chain and to perform a rotation from
the original model parameters to a set of principal axes based on the
preliminary Markov chain.  If the preliminary Markov chain were
representative of the full posterior distribution and it were possible
to find a linear transformation from the model parameters to an
efficient basis for taking steps in a Markov chain, then this would be
a good strategy.  We have attempted this method, but found less than
satisfactory results.  In particular, the length of the preliminary
Markov chain necessary for accurately estimating the best linear
transformation can be so long that we do not achieve our goal of
reducing the necessary computations.  This is related to the
observation that even the best linear combination of model parameters
often does not provide an efficient basis, due to the non-linear
nature of the Keplerian model.  For example, for planets with an
orbital period comparable to the duration of observations, there is
often a large non-linear degeneracy between the orbital period and
eccentricity.  Therefore, we found that a PCA-type analysis could be
useful for identifying and removing linear correlations, but it is not
sufficient for improving the efficiency of MCMC due to non-linear
correlations.  Instead we find that such PCA-type analyses are more
useful for providing intuition that can help guide the choice
of alternative CTPDFs.

\subsection{Alternative Choices for the Candidate Transition Probability Function}

We have developed a more general method of generating CTPDFs which is
capable of removing both linear and non-linear correlations.
One simple way of generating alternative CTPDFs is to identify
combinations of model parameters ($\vec{u}(\vec{\theta})$) which may
be better or more poorly constrained by the observational data than
the model parameters ($\vec{\theta}$).  A CTPDF can be constructed by
making a change of variables, taking a step which is Gaussian in the
transformed variables, and changing back to the original set of model
parameters.  By using the Metropolis-Hastings algorithm the Markov
chain will still be reversible and still converge to the same
distribution as if steps were taken in the original variables.  For
the acceptance probability in Eqn.\ \ref{AlphaEqn}, the function
$q(\vec{\theta}|\vec{\theta'})$ is proportional to $J = \left|
\partial \vec{u}(\vec{\theta})/\partial \vec{\theta} \right|$, the
magnitude of the determinant of the Jacobian of the transformation
between $\vec{\theta}$ and $u(\vec{\theta})$.
Below we present several examples of CTPDFs
which can speed convergence for some planetary systems and
observational data sets.

\subsection{Orbital Period and Phase}

The mean anomaly at time $t$ is given by
\begin{equation}
M(t) - M_o = \frac{2\pi}{P} \left( t - \tau \right)
\end{equation}
where $M_o$ is a constant model parameter, the orbital phase at
$t=\tau$.  Consider the case where the observational data constrain
the mean anomaly at $t=\tau'$ better than at $t=\tau$.  In that case,
a sensible choice for $\vec{u}$ might include the variable
\begin{equation}
M'_o \equiv M(\tau') =  M_o + \frac{2\pi}{P} \left( \tau' - \tau \right).
\label{eqnMt1}
\end{equation}
This generates an obvious CTPDF with the form of Eqn.\
\ref{eqnCandTransProb} using the variable $u_{M'_o}(\vec{\theta}) =
M'_o$. If $\vec{\theta}_{-M_o}$ (all the variables in the state
excluding $M_o$) are held constant, but $M_o$ is forced to change to
satisfy Eqn.\ \ref{eqnMt1}, then $M'_o$ is related to $M_o$ by a
simple linear shift.  A CTPDF based on this type of step in $M'_o$
will be just as efficient as one based on $M_o$.

The use of $M'_o$ can also lead to another new CTPDF involving the
orbital period.  We can take a step which alters both $M_o$ and $\log
P$ while holding $M'_o$ constant.  Due to the constraint, there is
still only one degree of freedom in this step, so we can still use
Eqn.\ \ref{eqnCandTransProb}, adding only one new step scale parameter
to $\vec{\beta}$.  The change of variable results in a Jacobian equal
to unity and hence the ratio
$q(\vec{\theta}|\vec{\theta}')/q(\vec{\theta}'|\vec{\theta}) = 1$ in
Eqn.\ \ref{AlphaEqn} for both these new step types.

Since we have assumed that the observations constrain $M'_o$ more
tightly than $M_o$, the pair $P$ and $M_o$ is more highly correlated
than the pair $P$ and $M'_o$.  The lower correlation will cause a
Markov chain to converge more quickly when taking steps in $\log P$
while holding $M'_o$ constant rather than taking steps in $\log P$
while holding $M_o$ constant.  To understand why taking steps in $\log
P$ with constant $M_o$ would slow the rate of convergence, let us
assume that the variances for the marginal posterior distributions of
$\log P$, $M'_o$ and $M_o$ are $\sigma^2_{\log P}$, $\sigma^2_{M'_o}$
and $\sigma^2_{M_o}$, with $\sigma^2_{M'_o} < \sigma^2_{M_o}$.  If a
step in $\log P$ is of size $\Delta \log P$, then the implicit step in
$M_o$ ($M'_o$) while holding $M'_o$ ($M_o$) constant will be of size
$\simeq 2\pi (\tau'-\tau) \Delta \log P / P$.  If $(\tau'-\tau)/P$ is
large, then the implicit change in the mean anomaly at $t=\tau$
($\tau'$) will often be larger than $\sigma_{M'_o}$, resulting in a
large fraction of trial states being rejected.  In order to maintain a
reasonable acceptance rate, the typical trial step sizes $\beta_{\log
P}$ will need to be reduced from $\simeq \sigma_{\log P}$ to $\simeq
\sigma_{M'} P \Delta \log P / (2\pi\Delta t)$.  However, the smaller
value of $\beta_{\log P}$ will mean that the Markov chain requires
more steps to explore the allowed range of orbital periods.  If
$\beta_{\log P} \ll \sigma_{\log P}$, then the Markov chain will
behave like a random walk, requiring a factor of $\sim (\sigma^2_{\log
P}/\sigma^2_{M'_o}) (4\pi^2 \Delta t^2/P^2)$ more steps to explore the
allowed range of orbital periods and initial mean anomalies by
stepping in $\log P$ while holding $M_o$ fixed than if stepping in
$\log P$ while holding $M'_o$ fixed.

In order to minimize the correlation between $\log P$ and $M(\tau')$
and to maximize the efficiency of our Markov chains, we want to choose
$\tau'$ to be the time at which the mean anomaly is best constrained.
Unfortunately, in practice we do not know the time at which $M(t)$ is
best constrained {\it a priori} and must make a reasonable
approximation.  Clearly, one should not hold $M(t)$ constant for a
time well before or after all the observations.  A much better initial
guess would be the weighted average of the observation times,
\begin{equation}
\label{Eqnt1}
\tau' = \frac{\sum_k t_k \sigma_k^{-2}}{\sum_k \sigma_k^{-2}}
\end{equation}
where $t_k$ is the time of the $k$th observation and $\sigma_k$ is the
uncertainty in the $k$th observation.  We have found this to be a good
initial guess for many practical radial velocity data sets.  If after
running a Markov chain it becomes clear that there is a significant
correlation between $\log P$ and $M_o$, then one could estimate a
better guess from the current Markov chain and start over with a new
choice for $\tau'$.

In this simple example, we could have simply chosen a different epoch
for defining our orbital parameters (i.e., set $\tau=\tau'$).  Indeed,
for the remainder of the paper, we will simply use $M_o$ assuming that
$\tau$ has been chosen wisely.  Nevertheless, we include this
methodical explanation to introduce our approach of using alternative
choices for CTPDFs.  Additionally, this complete development allows
for $\tau'$ to be a function of the model parameters at each state in
the Markov chain.  This may be useful for systems when we include a
noise source of unknown magnitude and the values of $\sigma_k$ depend
on the model parameter $\sigma_+$.  If the posterior distribution
permits a wide range of $\sigma_+$, then the time at which the mean
anomaly is best constrained may vary from model to model in a Markov
chain and it may be advantageous to recalculate $\tau'$ based on Eqn.\
\ref{Eqnt1} after each step in $\sigma_+$ to further improve the
efficiency of the Markov chain.  However, in our experience, the
practical differences are typically small for data sets where
$\sigma_+$ is small and the values $\sigma_{k,obs}$ are similar.

\subsection{$\omega+M_o$ and $\omega-M_o$}

For planets in low eccentricity orbits the longitude of pericenter can
be poorly constrained.  If the observations are consistent with $e=0$
and Markov chain takes steps in $e$ and $\omega$, then the Markov
chain would need to explore the entire range of $\omega$.  If
$\beta_{\omega}$ is small compared to $2\pi$ (e.g., due to correlation
with another model parameter), then the chain will be slow to
converge.  Since the initial mean anomaly is related to the time since
pericenter, the initial mean anomaly can only be well constrained when
the argument of pericenter is also well constrained.  When neither
$\omega$ or $M_o$ are well constrained individually, the observational
data can often better constrain $\omega+M(\tau)$ or $\omega+T(\tau)$,
the angular position of the star relative to the plane of the sky at
time, $\tau$.  The true anomaly $T$ is related to the mean anomaly via
Eqns.\ \ref{EccTrueEqn} \& \ref{KeplerEqn}.  While the relationship
between $M(t)$ and $T(t)$ is transcendental, we can approximate $T$
with a series expansion in $e$,
\begin{equation}
T \simeq M + 2e \sin M + \frac{5}{4} e^2 \sin 2M + O(e^3). 
\end{equation}
For low eccentricity systems, $T \simeq M$ and we find it advantageous
to use CTPDFs which are based on the variables $\omega+M_o$ and
$\omega-M_o$.  Clearly a prior which is uniform in the angles $\omega$
and $M_o$ is also uniform on the angles $\omega+M_o$ and $\omega-M_o$.
By using a transition probability function based on Eqn.\ \ref{eqnpdx}
and the variables $\omega+M_o$ and $\omega-M_o$, the Markov chain is
able to use a small scale, $\beta_{\omega+M_o}$, for steps in
$\omega+M_o$ and a larger scale, $\beta_{\omega-M_o}$, for steps in
$\omega-M_o$, since this choice of variables has a much lower
correlation.  We have found that this simple choice can greatly speed
convergence for low eccentricities and can also be useful for systems
with significant eccentricities.

\subsection{$e \sin \omega$ and $e \cos \omega$}

As discussed above, the argument of pericenter is typically not well
constrained for planets in low eccentricity orbits.  In Paper I, we
addressed this issue by using the variables $e \sin \omega$ and $e
\cos \omega$.  This works well for low eccentricity systems, since the
Markov chain can jump to arbitrary values of $\omega$ in a just one or
two steps when $e\simeq 0$.
One disadvantage of this approach is that a uniform prior in $e \sin
\omega$ and $e \cos \omega$ results in a non-uniform prior for $e$.
In Paper I a Markov chain was run with $p(e) = e$ and then importance
sampling was used to resample assuming a prior uniform in $e$.  This
typically worked well for low eccentricity systems.  However, it
required calculating longer Markov chains and checking that the
posterior distributions were sufficiently similar for importance
sampling to provide accurate results.  For some systems where the
eccentricity was poorly constrained, the method of Paper I resulted in
the Markov chain spending a larger than necessary amount of time
exploring high eccentricities and significantly reduced the efficiency
of the Markov chain.  Therefore, we now prefer to sample directly from
a prior that is uniform in $e$.  It is trivial to take steps in $e$
(holding $\omega$ constant) and $\omega$ (holding $e$ constant).

In order to explore parameter space quickly for low eccentricity
systems, we introduce two new CTPDFs based on Eqn.\
\ref{eqnCandTransProb} and the variables $e \sin \omega$ and $e \cos
\omega$.  We generate a trial state by taking a step in $e \sin
\omega$ and holding $e \cos\omega$ fixed or by taking a step in $e
\cos \omega$ and holding $e \sin \omega$ fixed.  The MH algorithm is
used to generate a reversible CTPDF and guarantee that the Markov
chain will still converge to the same $p(\vec{\theta}|\vec{d},
\mathcal{M})$ as it would if only taking steps in $e$ and $\omega$.

Given the domains of $e$ and $\omega$ ($0\le e \le 1$ and $0 \le
\omega \le 2 \pi$), it is easy to see geometrically that a uniform
prior in $e \sin \omega$ and $e \cos \omega$ would correspond to a
prior in $e$ and $\omega$ which is proportional to $e$.  Therefore,
when comparing the current state to a candidate state generated by
stepping in the variable $u_{e \sin \omega}$ or $u_{e \cos \omega}$,
the acceptance rate should decrease the probability of accepting the
model with the larger eccentricity.  More formally, Eqn.\
\ref{EqnHkToEomega} shows that
the the Jacobian of the transformation is $J = e$.  Therefore, when
taking steps in $e\cos\omega$ and $e\sin\omega$ rather than $e$ and
$\omega$, the ratio used in Eqn.\ \ref{AlphaEqn} of the MH algorithm
\be
\frac{q(\vec{\theta}|\vec{\theta}')}{q(\vec{\theta}'|\vec{\theta})} =
\frac{e}{e'}
\ee
is not a constant as it has been for previous CTPDFs.

Note that we have not yet completely specified the types of steps to
be taken.  Each time we change either $e \sin \omega$ or $e \cos
\omega$, we implicitly change other quantities, such as $e$, $\omega$,
and $\omega+M_o$.  If the quality of the orbital fit depends
sensitively on one of these quantities, then there is likely to be
another case of correlated variables and inefficient sampling of the
posterior.  Our simulations have shown that stepping in $e\sin\omega$
and $e\cos\omega$ while holding $P$, $K$, and $\omega+M_o$ fixed is
typically a good choice.

\subsection{$\omega\pm T_o$}

For systems with a significant eccentricity, $(T(t)-M(t))/2\pi\not\ll 1$,
so it may be preferable to take steps using the variables, $\omega+T(\tau)$
and $\omega-T(\tau)$ while holding the other fixed.  Differentiating
Eqns.\ \ref{EccTrueEqn} \& \ref{KeplerEqn}, we find 
\be
\frac{dT}{dM} =
\frac{\sqrt{1-e^2}}{(1-e\cos E)^2},
\ee
and the Jacobian of the
transformation from $\omega$ and $M_o$ to $\omega\pm T_o$ is 
$J = \left| 2\sqrt{1-e^2}(1-e\cos E)^{-2} \right|$.  
Again, we use a CTPDF based on Eqn.\ \ref{eqnpdx} with the acceptance probably in Eqn.\ \ref{AlphaEqn}
given by
\be
\frac{q(\vec{\theta}|\vec{\theta}')}{q(\vec{\theta}'|\vec{\theta})} = \frac{\sqrt{1-e^2} \left(1-e' \cos E'(\tau) \right)^2}{\sqrt{1-e'^2} \left(1-e \cos E(\tau) \right)^2}.
\ee
We find that the use of these steps can be useful for high eccentricity systems.

\subsection{Time of Pericenter}

When the time span of observations of an eccentric planet are
comparable to the orbital period, the time of pericenter may be more
tightly constrained by the observations than the initial mean anomaly.
Additionally, the correlation between the orbital period and initial
mean anomaly may still slow convergence.  By allowing steps in $T_p =
\tau - PM(\tau)/(2\pi)$ (holding $\log P$ fixed), it may be possible
to speed convergence.  The Jacobian of the transformation from $P$ and
$M_o$ to $P$ and $T_p$ is simply $J= \left|P/(2\pi)\right|$, so we use
a transition acceptance probability of
\begin{equation}
\frac{q(\vec{\theta}|\vec{\theta}')}{q(\vec{\theta}'|\vec{\theta})} = \frac{P}{P'}
\end{equation}
in Eqn.\ \ref{AlphaEqn} to generate a reversible CTPDF.

\subsection{$1/P$}

If there is a large range of possible orbital periods, then it can be
advantageous to take smaller steps when at short periods and larger
steps when at long periods.  While using $\log P$ results in a step
size scale that is proportional to the period, the scale for a change
in the orbital period to affect the velocities by a given amount
increases even faster, $\beta_{\log P}\sim P$.  Therefore, it can be
advantageous to take steps using Eqn.\ \ref{eqnCandTransProb} using
$u_{1/P}(\vec{\theta}) = 1/P$ with a step size scale, $\beta_{1/P}$.
Gregory (2005b) suggested a similar optimization, but did not give
details.  When replacing steps in $\log P$ with steps in $1/P$, the
acceptance probability for use in Eqn.\ \ref{AlphaEqn} is
\be
\frac{q(\vec{\theta}|\vec{\theta}')}{q(\vec{\theta}'|\vec{\theta})} = \frac{P'}{P}.
\ee
We find that this is particularly helpful for planets with a large
range of acceptable orbital periods.  This can occur for multimodal
posterior distributions (Gregory 2005a) or when the orbital period is
comparable to the duration of observations (Ford 2005a).

\subsection{$K\cos\omega$ \& $K\sin\omega$}

The radial velocity perturbation of a planet on a Keplerian orbit can
be written as a linear function of the variables $K\cos\omega$ and
$K\sin\omega$, if the orbital elements $P$, $e$, and $M_o$ are held
fixed.  This led us to consider CTPDFs based on the variables
$u_{K\cos\omega}(\vec{\theta}) = K\cos\omega$ and
$u_{K\sin\omega}(\vec{\theta}) = K\sin\omega$.  When replacing steps
in $\log K$ and $\omega$ with steps in $K\cos\omega$ and
$K\sin\omega$, the acceptance probability for use in Eqn.\
\ref{AlphaEqn} is
\be
\frac{q(\vec{\theta}|\vec{\theta}')}{q(\vec{\theta}'|\vec{\theta})} = \frac{K^2}{K'^2}.
\ee
Our tests found that these steps could be useful for long period planets.

\subsection{$\log \left(K\sqrt{1-e}\right)$ \& $\log \left(K/\sqrt{1-e}\right)$}

When a planet's orbital period is comparable to the duration of
observations, then there can be significant degeneracies between the
orbital parameters.  From studying the posterior distribution of
dozens of long period systems, we recognized that for such systems
$\log \left(K\sqrt{1-e}\right)$ and $\log \left(K/\sqrt{1-e}\right)$
often have lower correlations with the other orbital parameters than
$K$ and $e$ individually.  Therefore, we considered taking steps in
these variables.  When replacing steps in $\log K$ and $e$ with steps
in $\log \left(K\sqrt{1-e}\right)$ and $\log
\left(K/\sqrt{1-e}\right)$, the acceptance probability for use in
Eqn.\ \ref{AlphaEqn} is
\be
\frac{q(\vec{\theta}|\vec{\theta}')}{q(\vec{\theta}'|\vec{\theta})} = \frac{\left|1-e'\right|}{\left|1-e\right|}.
\ee
We found that taking steps in $K/\sqrt{1-e}$ while holding
$K\sqrt{1-e}$ could be useful for long period planes.

\subsection{Other Variables}

We have also considered taking steps in the numerous combinations of
variables $P (1-e)^{3/2} M_{*}^{1/2}$, $P^{2/3} (1-e) M_{*}^{1/3}$,
$P^{2/3} (1+e) M_{*}^{1/3}$, $K e \cos\omega$, $K\sqrt{1-e^2}$,
$K/\sqrt{1-e^2}$, $K\sqrt{1-e}$, and $\omega+2M_o$.  While some of
these have helped particular systems, we have not found these other
quantities to be advantageous in general.  Therefore, we suggest
monitoring the correlations among all model parameters,
$\vec{\theta}$, all variables being used for steps
$\vec{u}(\vec{\theta})$, and additional variables such as those above.
If statistics such as $\widehat{R}$ and $\widehat{T}$ indicate the
Markov chains are converging slowly for any of these variables, then
we suggest examining the correlations between all pairs of monitored
variables to see if additional CTPDFs may be useful for improving the
rate of convergence.

\subsection{Directly Sampling from Conditional Posterior Distribution}

For some variables it is possible to directly sample from the
conditional posterior distribution without resorting to
Metropolis-Hastings algorithm.  This reduces the autocorrelation of a
Markov chain, and hence increases the rate of convergence.
Unfortunately, if there are significant correlations between various
model parameters, then the Markov chain will still tend to have a
significant autocorrelation.

For example, the radial velocity predicted by the model is a linear
function of each of the constant velocity terms, $C_j$.  By assuming a
normal prior for $C$ and taking the limit as the variance goes to
infinity, we can compute the conditional posterior for the constant
terms assuming a uniform prior (Bullard, private communications).  We
find that the conditional posterior for the new constant terms $C_j$
given the data ($\vec{d}$) and the model parameters excluding $C_j$
($\vec{\theta}_{-C_j}$) is given by
$p(C_j | \vec{d}, \vec{\theta}_{-C_j},  \mathcal{M}) \sim 
 N \left( C'_j, \sigma_{C'_j}^2 \right)$
where 
\be
C'_j = \frac{\sum_k \left(d_k-v_{*,\vec{\theta}}(t_k, j_k | \vec{\theta}_{-C_j}, C_j=0,  \mathcal{M}) \right) \sigma_k^{-2} \delta_{j_k,j}}{\sum_k \sigma_k^{-2} \delta_{j_k,j}},
\ee
$\sigma_{C'_j}^2 = \left(\sum_k \sigma_k^{-2} \delta_{j_k,j} \right)^{-1}$,
and 
$\delta_{j_k,j}$ is the Kronecker delta.

\subsection{Numerical Tests}

We have constructed numerous simulated data sets for testing the
efficiency of Markov chains using various CTPDFs.  Here we concentrate
on quality data sets that provide significant constraints on the
orbital parameters.  We intentionally do not consider data sets where
a planet is only barely detected due to a small amplitude or a small
number of observations, as these can result in multi-modal posterior
distributions (Gregory 2005a) and even the optimizations presented in
this paper can struggle with multi-modal posterior distributions.  Our
simulated data sets assume $\simeq 80$ observations each with an
observational uncertainties of $\sigma_{k,obs}\simeq1$m s$^{-1}$.  We
include a planet causing a wobble of $K\simeq 50$m s$^{-1}$, typical
for an extrasolar giant planet.  We include jitter as a Gausian noise
source with $\sigma_+ = 2$m s$^{-1}$, so that
$K/\sqrt{\sigma_{k,obs}^2 + \sigma_+^2} \simeq 20$.  We consider
several orbital periods ranging from 1/30 - 1 times the time span of
the observations.  The observation times and uncertainties are based
on actual observation times and uncertainties from the California and
Carnegie planet search.  We then calculated many Markov chains for
each data set using various sets of CTPDFs to determine which result
in the most rapid rate of convergence.  We use these simulations to
compare the performance of various CTPDFs.  We present the median
stopping time for several choices of CTPDFs in Table \ref{Tab2}.  The
first two columns show the results for the CTPDFs very similar to
those used in Paper I.
\be
\vec{u}_{1}(\vec{\theta}) = ( \log P, \log K, e, \omega, M_o, C, \sigma_+ ),
\ee
\be
\vec{u}_{2}(\vec{\theta}) = ( \log P, \log K, e \sin (\omega), e \cos (\omega), M_o, C, \sigma_+ ),
\ee
The remaining columns show the results for some of the best CTPDFs
developed in this paper.  Note that the new CTPDFs often accelerate
convergence by one to nearly three orders of magnitude.  These CTPDFs
make it practical to perform Bayesian analyses of many single planet
systems with only $\sim10^4$ steps (using the strict stopping criteria
from \S\ref{StopCriteria}.  Even for a planet with an orbital period
equal to the duration of the observations and a large eccentricity, we
find stopping times of $\sim10^6$ steps.  Thus, these optimizations
presented in this paper, make it practical to perform Bayesian
analyses of the vast majority of extrasolar planets.  For reference,
we find that the computer time required is $\sim8\times10^{-7} N_{obs}
N_p L_c N_c$ seconds using a 3GHz Intel CPU, where $N_{obs}$ is the
number of observations, $N_p$ is the number of planets, $L_c$ is the
number of states in each chain, and $N_c$ is the number of chains, if
the planets are assumed to be on non-interacting Keplerian orbits.
While this study used $N_c=10$ chains to ensure accurate estimates for
the stopping time, experience has shown that $N_c\simeq5$ would be
sufficient for most applications.

\subsection{Recommendations}
\label{SRec}

We now offer some simple recommendations that can be generally
implemented.  We find that the optimal set of CTPDFs depends on the
types of orbital solutions consistent with the observational data.  In
particular, the optimal choice depends on the range of allowed
eccentricities.  Since the purpose of MCMC is to determine the allowed
range of orbital parameters, it may not be clear which set of CTPDFs
should be used.  In these cases, we suggest alternating between the
various types of CTPDFs suggested below.  While this is less efficient
than using only the CTPDF most appropriate for the particular planet,
it is also more robust.  Of course, preliminary analyses often provide
a good indication of whether the eccentricity is likely to be small or
large.  If it is clear that the eccentricity can not be large, then it
would be more efficient to use the CTPDFs suggested in \S4.13.1.

\subsubsection{Low Eccentricity Orbits}

For systems with low eccentricities, we find that taking steps in
the set of model parameters,
\be
\vec{u}_{3}(\vec{\theta}) = ( 1/P, \log K, e \sin (\omega), e \cos (\omega), \omega+M_o, C, \sigma_+ ),
\ee
results in good performance.  
Note that when we step in $e\sin(\omega)$ or
$e\cos(\omega)$, we hold $\omega+M_o$ constant.  The Jacobian of the
transformation relating $\vec{u}_{3}(\vec{\theta})$ to $\vec{\theta}$
is $J_1 = e/P$.
When taking Metropolis-Hasting steps, we use the ratio of the Jacobians of the two states 
\be
\frac{q(\vec{\theta}|\vec{\theta}')}{q(\vec{\theta}'|\vec{\theta})} = \frac{e P'}{e' P}
\ee
as the acceptance probability in Eqn.\ \ref{AlphaEqn} to ensure
that the Markov chains converge to the desired posterior distribution.  

Using CTPDFs based on $\vec{u}_{3}$, we found that Markov chains were
typically suitable for inference after $\sim10^{4-4.5}$ steps when
analyzing data for a planet with a moderate eccentricity ($e=0.5$) and
an orbital period longer than duration of observations.  For an
orbital period equal to the duration of observations, $\sim10^{4.7}$
steps were needed.  For very small eccentricities, $\sim10^{4.6-5.4}$
steps were needed.  Even for systems with large eccentricities,
$\sim10^{5-7}$ steps were needed.  These results indicate that this
set of CTPDFs is very efficient for nearly circular orbits and also
works across a broad range of eccentricities.  However, the reduction
in efficiency for high eccentricity and long period systems has
motivated us to test other sets of CTPDFs to further improve the
efficiency for such systems.

\subsubsection{Long Period \& High Eccentricity Orbits}
\label{SRecHiE}

For systems with high eccentricities, we find that alternating between
taking steps in $\vec{u}_{3}$ and two additions sets of model
parameters,
\bea
\vec{u}_{4} & = & \left( P^{-1}, K \sin (\omega), K \cos (\omega), e, \omega+T_o, C, \sigma_+ \right), \mathrm{and} \\
\vec{u}_{5} & = & \left( P^{-1}, \log\left[K\sqrt{1-e}\right], \log\left[P(1-e)^{3/2}\right], \omega, T_p, C, \sigma_+ \right)
\eea
is advantageous.  The Jacobians for the these transformation are $J_4
= \left|K^2\sqrt{1-e^2} P^{-1} \left(1-e\cos E(\tau)
\right)^{-2}\right|$ and $J_5 \sim 1/\left|1-e\right|$.  This
combination of CTPDFs provides a modest improvement for planets with a
moderate eccentricity, and is particularly useful for high
eccentricities.  Using this combination, the stopping times for high
eccentricity systems were reduced to $\sim10^{4.7-6}$ steps, provided
that orbital period is less than 80\% of the duration of observations.
Even for the most challenging cases considered ($e=0.8$ and $T_{obs}/P
= 1-1.25$, where $T_{obs}$ is the time span of observations), Markov
chains computed with these CTPDFs satisfied the stopping criteria from
\S\ref{StopCriteria} after $\sim10^{6.2-6.4}$ steps.

\section{Multiple Planet Systems}

So far the transition probability functions presented all involve the
orbital parameters of only a single planet.  Nevertheless, these
refinements are very valuable when applying MCMC to multiple planet
systems.  Since the correlations between orbital parameters for
different planets are often small, convergence towards the posterior
distribution of orbital parameters for the entire system can be
dramatically accelerated by using CTPDFs that allow the Markov chains
to explore the full range of allowed orbital parameters for each
planet efficiently.

\subsection{Monitoring Dynamically Significant Quantities}

It is likely that additional optimizations are possible for multiple
planet systems.  However the significant diversity of planetary
systems means that there is an enormous array of potentially useful
choices for the CTPDFs.  We hope that the methods explained in this
paper and the examples provided will make it straightforward for
researchers to construct transition probability functions which are
appropriate for the particular problem at hand and converge to the
desired posterior probability distribution.  Here, we point out a few
dynamically significant quantities that should at least be monitored
when testing the Markov chain for non-convergence.  In some cases,
these combinations may also be useful for including in the set of
variables used for CTPDFs.

\subsubsection{Secular Evolution}

For multiple planet systems dynamical stability may make it possible
to constrain the orbital inclination, $i$, and the longitude of
ascending node, $\Omega$ (see \S 5.2).  In particular, the secular
evolution of a multiple planet system is often qualitatively different
when the relative inclination, $i_{rel}$, is between $40^\circ$ and
$140^\circ$.  Indeed, often such large relative inclinations can be
excluded by imposing the requirement of long-term dynamical stability,
even if $\sin i\not\ll1$ for both planets.  It is important to
remember that the relative inclination of two planets (labeled in and
out) is not simply the difference of the inclinations to the line of
sight, but depends on the ascending nodes,
\be
i_{rel} = \cos i_{in} \cos i_{out} - \sin i_{in} \sin i_{out} \cos (\Omega_{in}-\Omega_{out}).
\ee
Depending on the regime, the secular inclination evolution can also be
sensitive to $\Delta\Omega = \Omega_{out}-\Omega_{in}$, $\Delta\varpi
= \varpi_{out}-\varpi_{in} = \Omega_{out} + \omega_{out}
-\Omega_{in}-\omega_{in}$, $e_{in}$, and $e_{in}/e_{out}$.

We recommend monitoring $\Delta\omega$ and $\log(e_{in}/e_{out})$ for
all multiple planet systems to ensure that any Markov chain used for
inference displays adequate mixing of these variables.  When including
additional constrains such as dynamical stability, we also recommend
monitoring $\Delta\Omega$, $\Delta\varpi$, and $i_{rel}$.  Such
monitoring is important for establishing the suitability of Markov
chains for inference about the secular evolution of multiple planet
systems, as done in Ford et al.\ (2005) where
$\omega_{in}-\omega_{out}$ and $\omega_{in} + \omega_{out}$ were also used
as the basis for CTPDFs.

\subsubsection{Evolution near a Mean Motion Resonance}

If the two planets planets are near a low order mean motion resonance,
then it is dynamically interesting to monitor the resonant variables.
In particular, it is of interest if these angles are librating or
circulating and, if librating, what their amplitude is.  For example,
in a nearly coplanar system with two planets near a $2:1$ mean-motion
resonance (e.g., GJ 876), the lowest order eccentricity mean motion
resonances variables are $\varphi_1 = \varpi_{in} + M_{in} - 2
M_{out}$ and $\varphi_2 = \varpi_{out} + M_{in} - 2 M_{out}$.  Since
these quantities vary over an orbital timescale, it is important to
monitor these variables at an epoch where they are relatively well
constrained.  If the system is librating, their amplitude of libration
also depends on $2P_{in}/P_{out}$.  Therefore, we recommend monitoring
the variables $\varphi_1$, $\varphi_2$, and $\log(2P_{in}/P_{out})$,
in addition to the variables related to the secular evolution of the
system.

\subsection{Dynamical Constraints \& Importance Sampling}

In multiple planet systems, the requirement that the planetary system
be dynamical stable (for some time, $t_{\max}$, which is less than or
equal to the age of the system) can place significant constraints on
orbital parameters.  Unfortunately, performing long-term n-body
integrations can require a substantial amount of computer time.  MCMC
can be used to improve the computational efficiency of dynamical
stability studies.  MCMC can be used to estimate a posterior
probability distribution for the orbital parameters which incorporates
only knowledge from the observations data (and not the requirement of
dynamical stability).  Then, that posterior distribution can be used
as the the initial conditions for the n-body integrations to test for
dynamical stability.  Initial conditions which result in dynamical
instability are rejected, while initial conditions which are found to
be stable for $t_{\max}$ are accepted for inclusion in a posterior
probability distribution which includes both the observational and
dynamical constraints.  This technique can be particularly useful for
efficiently constraining the inclination of a planet's orbit to the
line of sight, $i$, the ascending node, $\Omega$, and the relative
inclination of the orbital planets of two planets, $i_{rel}$.

Note that it is computationally advantageous to first calculate a
posterior distribution based on the observational data alone and use
this distribution as the initial conditions for a series of
integrations to test for their long-term stability (requiring
$\sim10^{4-10}$yr integrations).  When calculating the posterior
distribution based on the observational data alone, the integrations
only need to last for the duration of observations (e.g.,
$\sim1-10$yr).  When testing the long-term stability, much longer
integrations are typically required (e.g., $\sim10^{4-10}$yr).
Therefore, separating these two dramatically reduces the total length
of integrations necessary, and the final estimate of the posterior
distribution (including both observational and dynamical constraints)
is still accurate, provided that there is significant overlap between
the posterior based on observations only and the set of solutions
satisfying the dynamical constraints.  An additional improvement in
computational efficiency is possible, since the length of the Markov
chain for calculating the posterior based solely on observational data
can be orders of magnitude greater than the number of initial
conditions for the long integrations to test dynamical stability.  By
selecting a small subset of states from a much longer Markov chain, we
can virtually eliminate the correlation between the sets of initial
conditions used for long-term integrations, reducing the number of
long integrations necessary to accurately characterize the long-term
stability.

In many multiple planet systems, the mutual planetary perturbations
are negligible over the timescale of the observational data, even if
they may be significant on secular timescales (e.g., $\upsilon$
Andromedae, 47 UMa).  In these cases, the Markov chains described in
\S 5.2 can be calculated assuming that each planet is on an
independent Keplerian orbit.  When this is a good approximation the
necessary computation time is greatly reduced, as the direct
integration of the few-body problem for the duration of observations
can be replaced by solving the Kepler equation for each planet at each
observation time.

We can further improve the accuracy of this approximation by first
calculating the posterior distribution for model parameters assuming a
likelihood function based on the approximation of independent
Keplerian orbits.  A subset of the states from the Markov chains using
the approximate model can then be used as the initial conditions for
direct n-body integrations, and importance sampling can be used to
calculate weights for each set of initial conditions.  It is possible
to obtain a sample from the actual posterior distribution (based on
the observations and direct n-body integrations) using these initial
conditions, weights, and rejection sampling.  This sample can then be
used as the initial conditions for long-term integrations to impose
the requirement of long-term dynamical stability.

One concern with this method is that if the observations constrain
parameters more tightly in the approximate model than in the full
model, then this two-stage method could underestimate the uncertainty
of the parameters in the full model.  This can be avoided by altering
the target distribution for on the approximate model, so that it is
broader than the posterior distribution based on the full model.  We
propose two ways to make such a modification.  Instead of initially
calculating Markov chains with a target distribution equal to the
posterior of the approximate model, $p(\vec{\theta}|\vec{d},
\mathcal{M}) \sim p(\vec{\theta}, \mathcal{M}) p(\vec{d}|\vec{\theta},
\mathcal{M})$, we can calculate Markov chains with a target
distribution equal to $p(\vec{\theta}|\vec{d}, \mathcal{M}) \sim
p(\vec{\theta}) p(\vec{d}|\vec{\theta}, \mathcal{M})^\gamma$.  By
setting $\gamma<1$, we can choose the target distribution to be
broader than the posterior distribution in the approximate model.
An alternative method is to artificially increase the estimates of the measurement errors when calculating the approximate model, $\sigma^2(t_i) \leftarrow \sigma^2(t_i) +
\sigma_a^2$.  We prefer this method when a good estimate for
$\sigma_a^2$ is available, so the value for $\sigma_a^2$ can be chosen
based on the magnitude of the expected deviations from independent
Keplerian orbits.  This method was applied and tested for the
analysis of the $\upsilon$ Andromedae system in Ford et al.\ (2005).

\section{Example Applications}

In this section we demonstrate the improvements in efficiency of MCMC
on several actual extrasolar planet systems.  We include one
particularly interesting short-period planet, one recently announced
long-period planet, one slowly interacting multiple planet system, and
one rapidly interacting multiple planet system.

\subsection{HD 209458}

\begin{figure}[ht]
\plotone{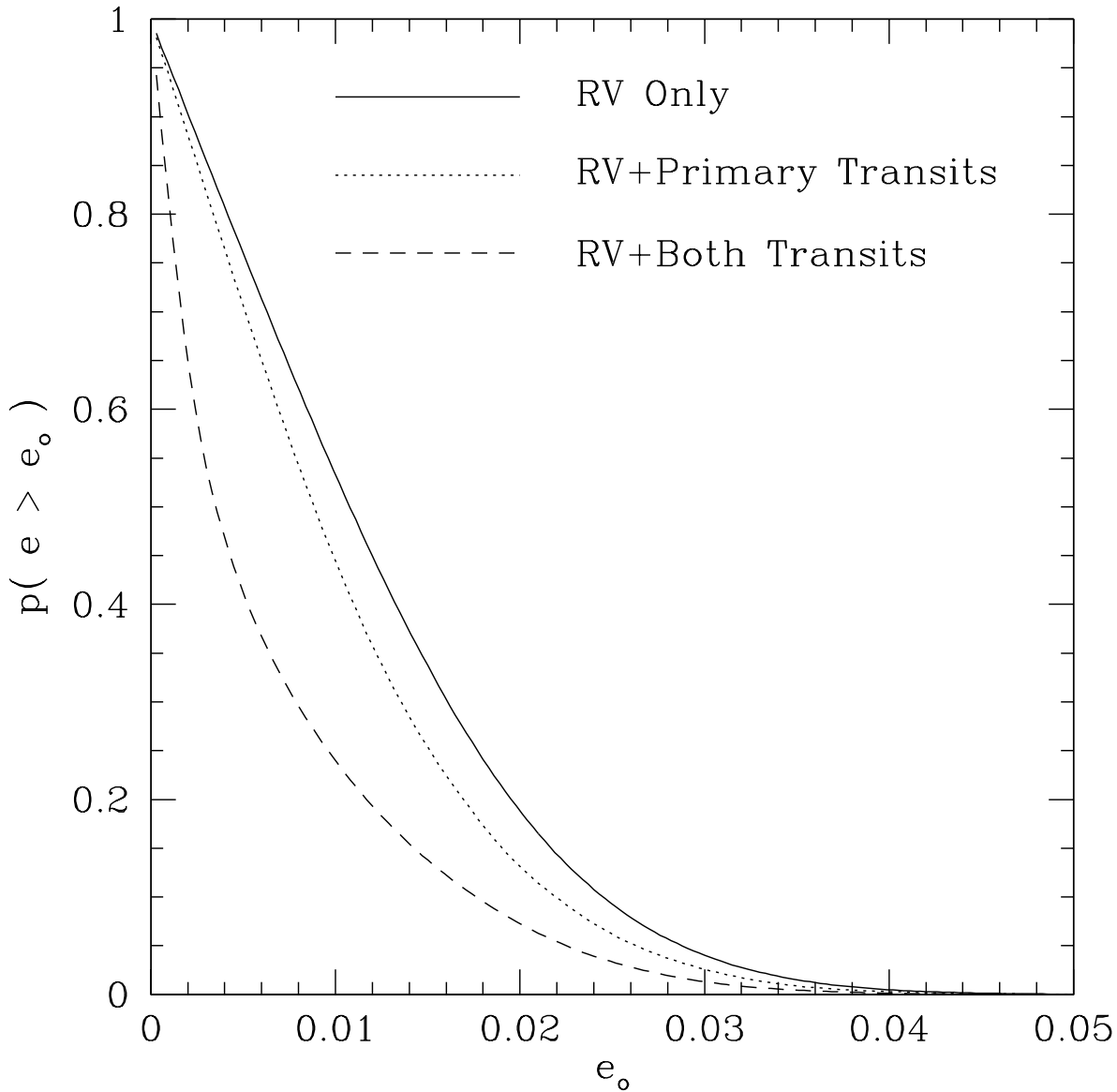}
\caption[hd209458_e.eps]{
\noindent
The posterior probability that the eccentricity of HD 209458b is
greater than $e_o$.  The solid curve is based on the radial velocity
observations only.  The dotted curve assumes a fixed orbital period
and phase from the optical primary transit observations.  The dashed
curve also incorporates the time of the infrared secondary transit.
\label{HD209458}}
\end{figure}

The transiting planet around HD 209458 has provoked considerable
interest on account of its large radius (Brown et al. 2001).
Bodenheimer, Laughlin \& Lin (2003) offered a potential explanation,
namely that the radius could be inflated due to tidal heating if the
planet still had an eccentricity of $\simeq0.03$.  Additional radial
velocity observations allowed radial velocities to impose a stronger
constraint on the allowed range of eccentricities (Laughlin et al.\
2005).  In Fig.\ \ref{HD209458} we show the results of a Bayesian
analysis of the orbit of HD 209458b.  The solid curve is based solely
on the radial velocity observations using the MCMC techniques
described in this paper.  The dotted curve is based on simulations in
which the orbital period and phase have been fixed based on
observations of the primary transits (Wittenmyer 2004; Laughlin et
al.\ 2005).  The dashed curve also incorporates the observation of the
secondary eclipse observed in the infrared.  To compute this last
cumulative posterior distribution, we use MCMC to sample from the
posterior distribution for orbital parameters based on the radial
velocity and primary eclipse observations.  We then treat this as the
prior distribution for predicting the midtime of the secondary eclipse
observed on December 6/7, 2004.  We use a Gaussian likelihood with a
standard deviation of 7 minutes and zero mean offset from the time
predicted for a circular orbit (Deming et al.\ 2005).  We then
calculate the marginal posterior distribution for the eccentricity and
integrate to obtain $p(e>e_o)$, the probability that the eccentricity
of HD 209458b is at least $e_o$.  Based on this posterior probability
distribution, we find that $p(e>0.008) = 0.32$, $p(e>0.023) = 0.05$,
and $p(e>0.042) = 0.001$.  Thus, it is unlikely that the current
eccentricity of HD 209458b is sufficient to explain the unexpectedly
large radius of the planet.

\subsection{HD 117207}

Next, we consider the recently announced long period planet orbiting
HD 117207.  Radial velocity observations spanning 7 years were used to
derive an orbital period of $7.2\pm0.3$ years and an eccentricity of
$0.16\pm0.05$ (Marcy et al.\ 2005), where the error estimates were
based on bootstrap-style resampling.  The residuals to the best-fit
orbit are consistent with a combination of measurement uncertainties
(typically $\simeq 3$m s$^{-1}$) and the expected stellar jitter based
on Ca II H \& K emission, $3.6$m s$^{-1}$.  As noted by Marcy et al.\
(2005) the observations provide only modest constraints on the orbital
parameters, since the orbital period is comparable to the duration of
observations.

We use MCMC to perform a Bayesian analysis of the orbital parameters
for this system, using the CTPDFs optimized for such systems from
\S\ref{SRecHiE} (see Fig.\ 2).  Our Bayesian analysis shows that the
$68\%$ credible interval for the marginalized posterior distribution
includes periods in the range 7.3-11.7 years and eccentricities from
0.11 to 0.29.  Thus, the uncertainties in the orbital parameters are
significantly larger than suggested by bootstrap-style resampling.  We
note that even the constraints from our analysis may be optimistic,
since we assume that there is only one planet in the system.  If we
also consider models which allow an additional planet, then the
posterior distribution becomes even more broad.

\subsection{$\upsilon$ Andromedae}

Many of the optimizations to the MCMC algorithm described in this
paper have already been applied to the three planet system around
$\upsilon$ Andromedae (Ford et al.\ 2005).  They used MCMC and several
of CTPDFs from \S4 to sample from the posterior probability
distribution for orbital parameters assuming independent Keplerian
orbits.  That sample was used as the initial conditions for n-body
integrations of the three planet system.  N-body integration also
demonstrated the robustness of these constraints to including the
mutual planetary perturbations.  Long-term direct n-body integrations
were then used to place additional constraints on the orbital
parameters, including the inclinations of the orbits to the plane of
the sky and especially the relative inclinations between the orbits,
by demanding dynamical stability.  The resulting distribution for the
dynamical state of the $\upsilon$ Andromedae system provided strong
evidence that the current eccentricities of planets c \& d are the
result of long-term mutual planetary interactions after planet d
received an impulsive perturbation (most likely due to another unseen
planet; Ford et al.\ 2005).

\subsection{HD 37124}

Radial velocity observations of the star HD 37124 have revealed three
planets (Vogt et al.\ 2005).  However, the discovery paper gave two
sets of orbital solutions with very different properties for planet d:
a preferred solution for planet d with an orbital period near 2200d,
and an alternative solution with planet d in a $\sim30$d orbital
period.

In principle, one could perform a Bayesian analysis of all three
planets using MCMC simulations.  In practice, this is not practical
with the numerical techniques described in this paper.  In particular,
the CDPDFs used in this paper allow only local jumps based on the
current state, and they are not appropriate for posterior PDFs which
have multiple peaks separated by very deep and wide valleys.  In the
case of HD 37124, we are particularly interested in the relative
probability of two orbital solutions with very different periods
($\sim30$d and $\sim2200$d), so it is important to use a numerical
technique which is capable of handling such strongly multi-modal
posterior PDFs.  One possibility is the technique of parallel
tempering (e.g., Gregory 2005).  Here we present an alternative method
that is computationally more efficient and more robust for identifying
small peaks in the posterior PDF.  This is possible because the method
described below assumes that the orbital parameters for the first two
planets are not correlated with the orbital parameters of the third
planet.

Since the orbits of two of the planets are well constrained by
observations, we sample from the posterior distribution using a model
that includes only the two previously discovered planets (b \& c)
modeled as on independent Keplerian orbits.  We use this sample to
calculate the posterior predictive distributions at the times of the
actual observations and subtract the observed velocities to obtain the
posterior predictive residual distribution at each time.  We then
analyze these velocity assuming they are due to a third planet (d),
also assumed to be on an independent Keplerian orbit.  We calculate
the posterior distribution marginalized over all model parameters
except period to estimate the significance of the detection of the
third planet.  This allows us to quantify the relative probabilities
of these two solutions for the third planet.  We find that $\go99\%$
of the posterior probability distribution occurs near the orbital
solution with an $P_d\sim2200$d.

While the above technique is very useful for searching for possible
alternative orbital solutions for planet d, it could underestimate the
uncertainty in model parameters if the orbital parameters for planets
b \& c are correlated with the orbital parameters for planet d.
Therefore, after we identified a single plausible orbital period, we
perform MCMC simulations using a three planet model to determine the
posterior PDFs for the orbital parameters.  Unfortunately, the
discovery paper provided no estimates of the uncertainties in model
parameters.  Therefore, we present marginalized posterior probability
distributions for the model parameter based on three planet MCMC
simulations near the preferred solution (Fig.\ 3).

We have used this sample to calculate the posterior predictive
distribution for the velocity as a function of time for several years
into the future (Fig.\ 4).  The width of the predictive distribution
is presently $\simeq5$m s$^{-1}$, but increases to over $50$m s$^{-1}$
within a year.  This is not simply a result of gradual loss of
information with time, as the width of the predictive distribution
increases and decreases significantly in a quasi-periodic manner.
Radial velocity observations taken at the times when the predictive
distribution has an unusually large width would be particularly
valuable (Loredo 2003; Ford 2005b).  It is important to note that the
suggested times are not merely due to aliases when the star is behind
the sun, below the horizon from Keck, or during new moon (when radial
velocity planet surveys are unlikely to have telescope time), as the
peaks in the width of the predictive distribution include times when
HD 37124 is observable from Keck and during full moon.  Therefore,
there is no practical barrier that prevents measuring velocities near
the times when the width of the posterior predictive distribution is
maximized.

\subsection{GJ 876}
\label{SGJ876}

The multiple planet system around the M dwarf GJ 876 is particularly
interesting, as there are two massive planets in a 2:1 mean motion
resonance (Marcy et al.\ 2001).  The mutual planetary perturbations
are sufficiently strong that the orbits are precessing at
$\sim40^\circ$/yr as shown in Paper I.  Additional observations have
been combined with direct n-body integrations to further constrain the
allowed masses and orbital parameters and to detect a very low mass
planet in a short period orbit (Rivera et al.\ 2005).  Here we perform
a Bayesian analysis of this system using a simplified model of
precessing Keplerian orbits (Paper I).  This model captures the most
important aspects of the interaction between the outer two planets by
adding a single free parameter ($\dot{\omega}_{bc}$, the precession
rate of planets b \& c), whereas a full dynamical model requires five
additional free parameters ($i$ and $\Omega$ for each of the three
planets, minus one parameter which merely describes the rotation of
the entire system).  If we assume that the orbital planes are
coplanar, then the physical system has one degree of freedom (the
inclination of the orbital plane to the sky) which is not included in
our model, and our model includes one parameter (the precession rate)
which is really a function of the other model parameters.  Therefore,
in principle for each set of model parameters, we could determine the
inclination at which the physical precession rate would equal the
precession rate in the model.  Unfortunately, the precession rate of
the outer two planets is a non-linear function of multiple model
parameters (inclinations and eccentricities) each of which have
significant uncertainties, so that only loose constraints can be
placed on each of $i_b$, $\Omega_b$, $i_c$, and $\Omega_c$.
Therefore, we believe it makes sense to report the measurement of
$\dot{\omega}_{bc}$, rather than several model parameters which are
not yet well constrained.

We have used methods similar to those described in the previous
section to analyze the HD 37124 system.  Our analysis confirms the
reality of a periodic signal with period, $1.9378$d (see Fig.\ 5), as
reported by Rivera et al.\ (2005).  We also see smaller peaks near the
periods of 0.66d, 2.05d, and 0.40d, but the posterior probability of
all of these peaks sum to less than 0.1\% of the primary peak.  Note,
that our analysis reveals no significant peak near the periods of 9,
13, or 120d, which were mentioned by River et al.\ (2005).

While our analysis is based on a simpler dynamical model than that of
Rivera et al. (2005), our method has several advantages.  First, we
use Bayesian rather than maximum likelihood methods, which allows us
to make quantitative statements about the probability for secondary
peaks.  Second, our model captures the important planetary interaction
by adding only a single free parameter, and yet our best-fit model
provides a slightly smaller $\sqrt{\chi^2_\nu}$, $\chi^2$, and rms
velocity scatter than any of the full n-body models reported in Rivera
et al. (2005).  This is because the overall precession of the outer
two planets is well constrained by the observational data, but the
data only weakly constrain the amplitude of the other mutual planetary
perturbations.  Additionally, our use of the posterior predictive
residuals rather than the best-fit residuals provides a more robust
tool when searching for additional periodicities.  Finally, a secure
detection of planet d with our simple model alleviates concerns that
the periodicity might have been introduced by the dynamical modeling.

It should be cautioned that while our analysis demonstrates the
reality of a 2.9d period in the velocity residuals (after subtracting
the velocity perturbations by outer two planets), it does not
necessarily imply that the signal is due to a planet.  For example, in
principle, a large moon around one of the outer two planets could also
cause a periodic velocity perturbation.  In the case of GJ 876, we
have verified that the amplitude of the 2.9d signal is too large to be
caused by a moon around either of the outer two planets.
Nevertheless, as radial velocity searches detect periodicities with
decreasing amplitudes, one should consider all possible causes for the
detected velocity perturbations.

\section{Conclusions}

This and previous papers have demonstrated numerous advantages using
Bayesian inference for analyzing the orbital parameters of extrasolar
planets.  Unfortunately, Bayesian analyses typically require
calculating multidimensional integrals which can be computationally
challenging.  In Paper I, we introduced the method of MCMC, the Gibbs
sampler, and the MH algorithm for performing these integrals.  This
made it practical to use MCMC to perform Bayesian analyses of the
orbital parameters of several extrasolar planets.  In this paper, we
developed and tested several alternative CTPDFs to further improve the
efficiency of MCMC.  In particular, we recommend two sets of CTPDFs
which together can accelerate the convergence of Markov chains by
multiple orders of magnitude (see \S\ref{SRec}).  These improvements
make it practical to analyze all the known extrasolar planets and even
multiple planet systems.  We demonstrate our optimized MCMC algorithms
by analyzing several actual extrasolar planet systems.  We anticipate
applying these algorithms to a large number of extrasolar planet
systems to obtain more accurate orbital parameters and improve
dynamical investigations of multiple planet systems.

Several significant challenges still prevent fully Bayesian analyses
from being routinely applied to extrasolar planet observations.  In
particular, it can be quite computationally demanding to sample from
the posterior distribution when the posterior distribution has several
well separated modes.  This commonly occurs for low-mass planets that
are not yet clearly detected.  It would be desirable to be able to
analyze such planets, since the marginally detected planets can
influence the orbital parameters derived for other planets that have
already been clearly detected around the same star.  Another important
challenge is developing computationally efficient means of performing
Bayesian model selection, i.e., simultaneously considering models with
zero, one, two, or more planets.  This would allow planet detections
to be based on Bayesian rather than maximum likelihood analyses.
Gregory (2005a) has suggested parallel tempering as one possible
approach, however even more efficient algorithms would be desirable.

\acknowledgments We thank Floyd Bullard, Peter Driscol, Debra Fischer,
Greg Laughlin, Geoff Marcy, John Rice, Scott Tremaine, and an
anonymous referee for valuable discussions.  This research was
supported in part by NASA grant NNG04H44g, and the Miller Institute
for Basic Research.  This research used computational facilities
supported by NSF grant AST-0216105.

%


\clearpage

\begin{deluxetable}{ll}
\tablewidth{0pt}
\tablecaption{Symbols Used in Multiple Sections
\label{Tab1}}
\tablehead{\colhead{ Symbol } 	 & \colhead{ Explanation } } 
\tabletypesize{\scriptsize}
\startdata
\cutinhead{ Indexes }  \\ 
$\mu$ & indexes functions of model parameters \\
$\nu$ & indexes model parameters \\
$i$ & indexes state in Markov chain \\
$j_k$ & indexes observatory for $k$th observation \\
$p$ & indexes planets \\
$k$ & indexes observations \\
\cutinhead{ Variables }  \\ 
%
$\vec{\beta}$ & vector of scale parameters \\
$\chi^2$ & chi-squared statistic \\
$\omega$ & argument of periastron \\
$\Omega$ & longitude of ascending node \\
$\sigma_{k}$ & effective uncertainty for $k$th observation \\
$\sigma_{k,\mathrm{obs}}$ & observational uncertainty in $k$th observation \\
$\sigma_{+}^2$ & variance due to stellar jitter \& unseen planets \\
$\tau$ & time of initial epoch \\
$\tau'$ & time of alternative epoch \\
$\vec{\theta}$ & model parameters \\
$\vec{\theta}'$ & model parameters for trial state \\
$\vec{\theta}_i$ & model parameters for $i$th state of Markov chain \\
$\vec{\theta}_p$ & set of model parameters for planet $p$ \\
$\upsilon$ & degrees of freedom for Student's $t$-distribution \\
%
$C_j$ & constant velocity offset for $j$th observatory \\
$d_k$ & value of $k$th observation \\
$d_{k,\theta}$ & prediction for $k$th observation with model, $\vec{\theta}$\\
$e$ & orbital eccentricity \\
$i$ & inclination of orbital planet relative to plane of sky \\
$J$ & Jacobian of a transformation \\
$K$ & velocity semi-amplitude \\
$m$ & planet mass \\
$m_{\min}$ & minimum planet mass ($\sin i=1$)\\
$M_*$ & stellar mass \\
$M$ & mean anomaly \\
$M_o$ & mean anomaly at epoch $t=\tau$ \\
$M'_o$ & mean anomaly at epoch $t=\tau'$ \\
$\mathcal{M}$ & model description \\
$P$ & orbital period \\
$t_k$ & time of $k$th observation \\
$T$ & true anomaly \\
$T_{obs}$ & time span of observational data \\
$T_p$ & time of pericenter \\
$v_{*,\mathrm{obs}}(t,j)$ & observed radial velocity at time $t$ \& observatory $j$ \\
%
\cutinhead{ Functions }  \\ 
%

$p(\vec{d},\vec{\theta} | \mathcal{M})$ & joint PDF for data \& model parameters in model $\mathcal{M}$\\
$p(\vec{d} | \vec{\theta}, \mathcal{M})$ & likelihood assuming model $\mathcal{M}$ \\
$p(\vec{\theta} | \vec{d} , \mathcal{M})$ & posterior PDF assuming model $\mathcal{M}$ \\
$\widehat{R}(z)$ & estimated Gelman-Rubin statistic for $z$ \\
$\widehat{T}(z)$ & estimated effective independent draws for $z$ \\
$u_{\mu}(\vec{\theta})$ & a function of model parameters \\
%
\enddata
\end{deluxetable}

\begin{deluxetable}{lllll}
\tablewidth{0pt}
\tablecaption{Median $\log_{10} N_{\mathrm Stop}$
\label{Tab2}}
\tabletypesize{\scriptsize}

\tablehead{\colhead{$T_{obs}/P$} 	 & \colhead{ $u_1$\tablenotemark{a}} & \colhead{ $u_2$\tablenotemark{b}} & \colhead{ $u_3$\tablenotemark{c}} & \colhead{$(u_3,u_4,u_5)$\tablenotemark{d}} } 
\startdata
\cutinhead{ e = 0.01 }  \\ 
1 	 & 7.8 & 7.9 & 5.4 & 5.5  \\ 
1.25 	 & 7.7 & 7.7 & 5.2 & 5.3  \\ 
1.5 	 & 7.7 & 7.7 & 4.6 & 5.2  \\ 
1.75 	 & 7.8 & 7.8 & 5.3 & 5.4  \\ 
2 	 & 7.9 & 7.9 & 4.9 & 5.0  \\ 
3 	 & 7.7 & 7.9 & 5.0 & 5.2  \\ 
10 	 & 7.8 & 7.9 & 4.8 & 5.2  \\ 
30 	 & 7.8 & 7.8 & 5.2 & 5.2  \\ 
\cutinhead{ e = 0.1 }  \\ 
1 	 & 6.2 & 6.3 & 4.7 & 4.7  \\ 
1.25 	 & 6.1 & 6.0 & 4.2 & 4.2  \\ 
1.5 	 & 5.9 & 5.9 & 4.1 & 4.2  \\ 
1.75 	 & 5.8 & 5.9 & 4.2 & 4.2  \\ 
2 	 & 6.0 & 6.0 & 4.0 & 4.1  \\ 
3 	 & 5.9 & 5.8 & 4.0 & 4.1  \\ 
10 	 & 5.9 & 5.8 & 4.0 & 4.1  \\ 
30 	 & 5.9 & 5.8 & 3.9 & 4.0  \\ 
\cutinhead{ e = 0.5 }  \\ 
1 	 & 5.3 & 5.3 & 4.7 & 4.7  \\ 
1.25 	 & 4.9 & 4.8 & 4.5 & 4.5  \\ 
1.5 	 & 4.7 & 4.7 & 4.4 & 4.3  \\ 
1.75 	 & 4.7 & 4.8 & 4.4 & 4.3  \\ 
2 	 & 4.8 & 4.9 & 4.5 & 4.3  \\ 
3 	 & 4.7 & 4.7 & 4.4 & 4.4  \\ 
10 	 & 4.8 & 4.8 & 4.3 & 4.4  \\ 
30 	 & 4.7 & 4.6 & 4.2 & 4.2  \\ 
\cutinhead{ e = 0.8 }  \\ 
1 	 & 7.2 & 7.1 & 7.0 & 6.2  \\ 
1.25 	 & 7.3 & 7.5 & 7.7 & 6.4  \\ 
1.5 	 & 7.0 & 6.6 & 7.1 & 6.0  \\ 
1.75 	 & 5.4 & 5.6 & 5.9 & 5.2  \\ 
2 	 & 5.5 & 5.5 & 5.8 & 4.9  \\ 
3 	 & 5.6 & 5.5 & 5.8 & 5.4  \\ 
10 	 & 5.8 & 6.2 & 6.1 & 5.5  \\ 
30 	 & 5.7 & 6.2 & 6.3 & 4.7  \\ 
\enddata
\tablecomments{We list the median stopping time for Markov chains computed using four different sets of CTPDFs.  The collums labled $u_1$ and $u_2$ show stopping times using CTPDFs from Paper I.  The collums labled $u_3$ and $(u_3,u_4,u_5)$ are based on the CTPDFs recommended in \S\ref{SRec}.}
\end{deluxetable}

\clearpage

\begin{figure}[ht]
\plotone{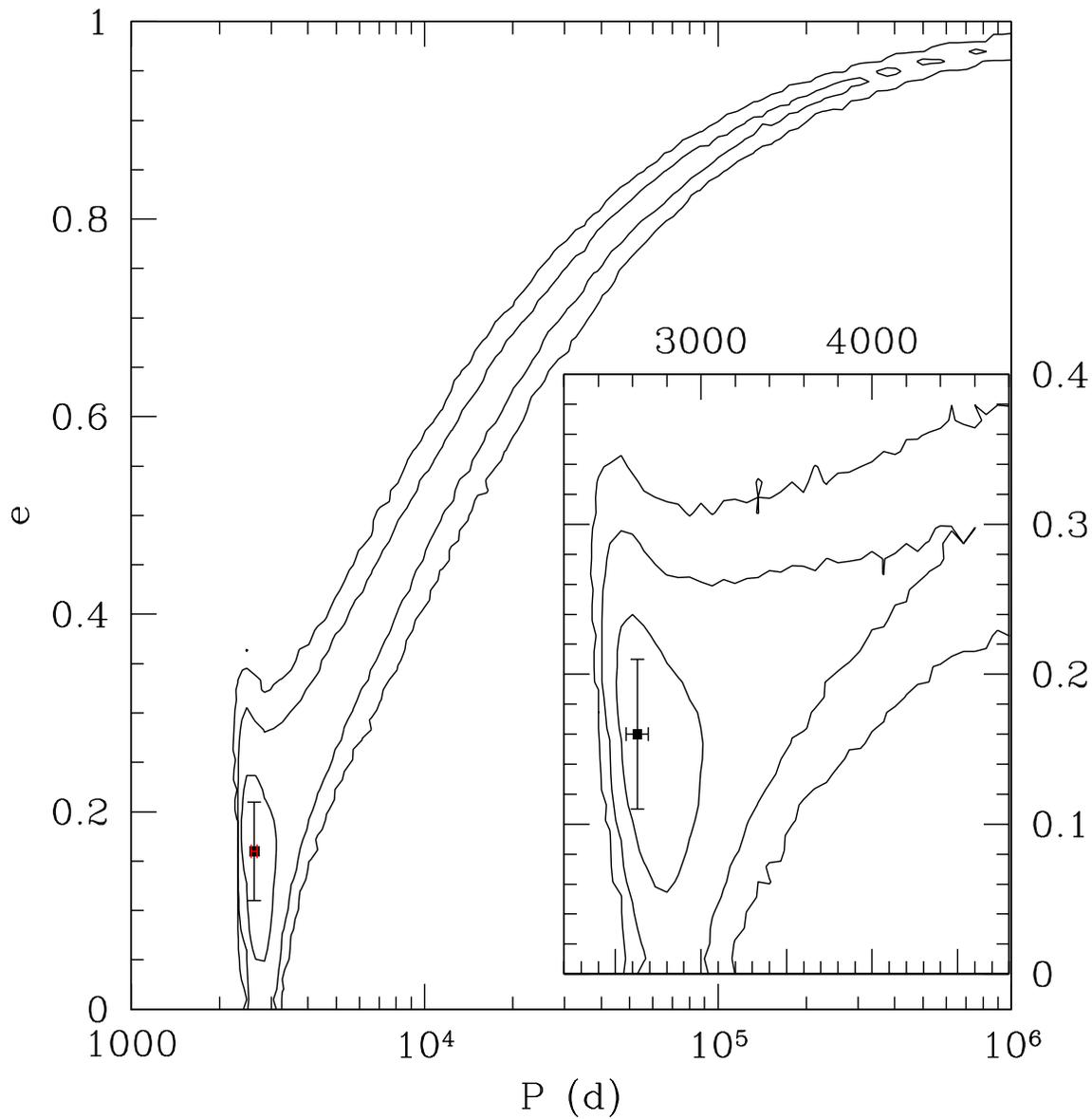} 
\caption[hd117207.eps]{
\noindent
The marginalized posterior probability for the period and eccentricity
of HD 117207b.  The contours indicate 1, 2, and 3$-\sigma$ credible
intervals (defined to contain 68.3\%, 95.4\%, and 99.73\% of the
posterior probability distribution).  The point and error bars show
the published period, eccentricity, and error estimates (Marcy et
al. 2005).
\label{HD117207}}
\end{figure}

\begin{figure}[ht]
\plotone{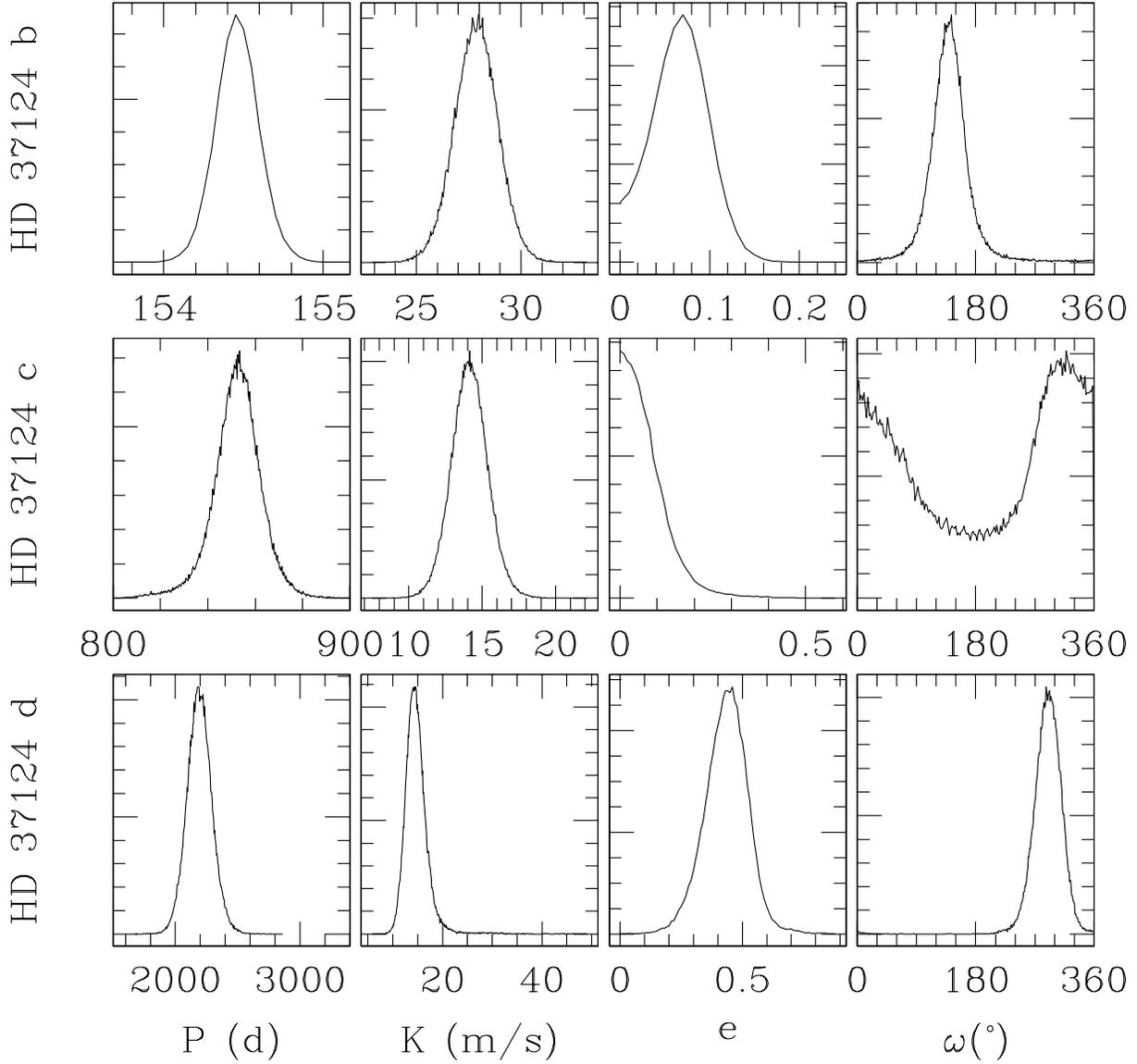} 
\caption[hd37124mpdf.eps]{
\noindent
The marginalized posterior probability density for the orbital
parameters of each of the three planets in the HD 37124 system.  These
distributions are based on a three planet model with no mutual
interactions.  
\label{hd37124mpdf}}
\end{figure}

\begin{figure}[ht]
\plotone{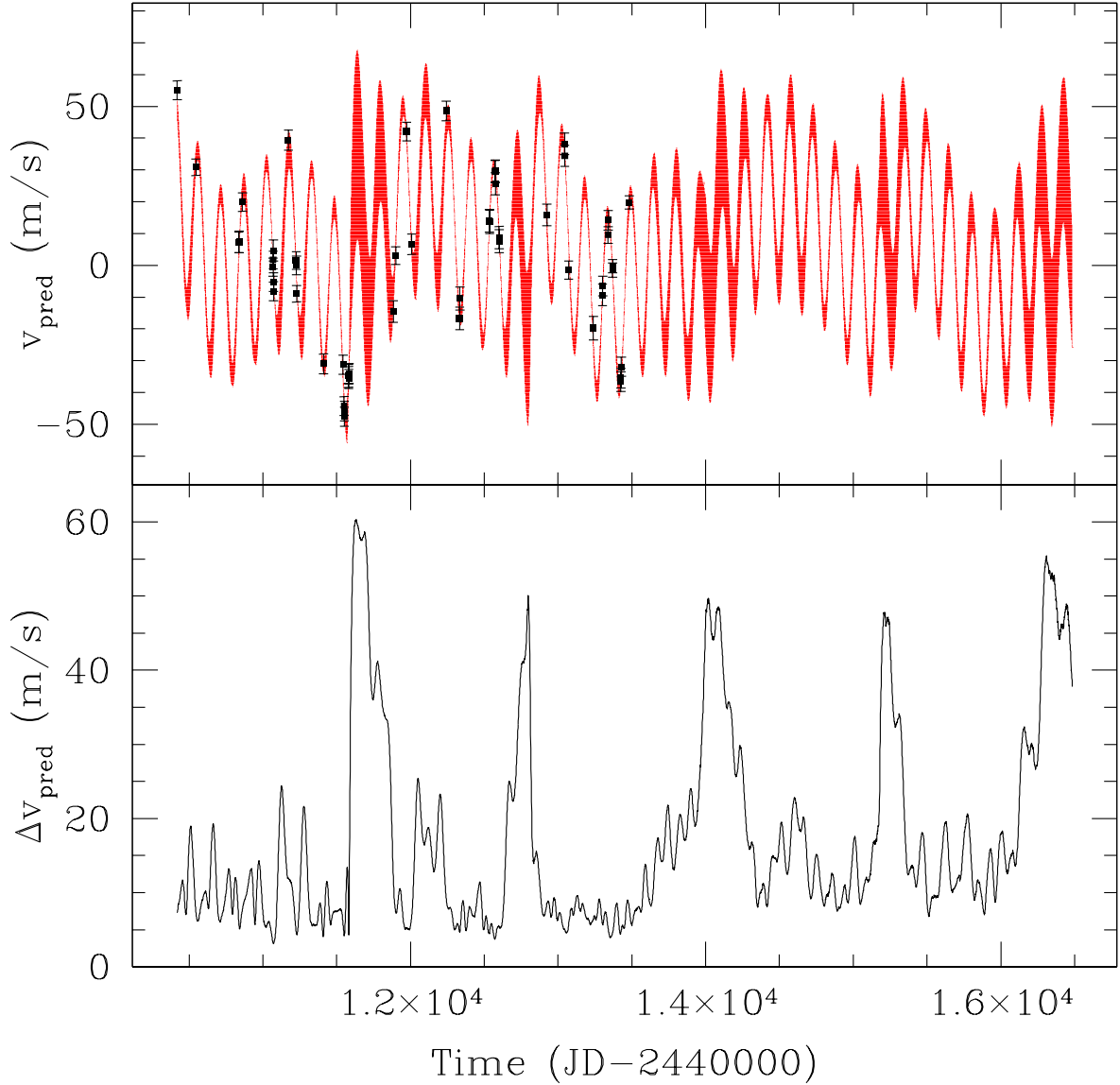} 
\caption[hd37124ppdf.eps]{
\noindent
The shaded region in the top panel shows the 68\% credible interval
for the posterior predictive distribution for the velocity of HD 37124
as a function of time.  We calculate the distribution of the predicted
velocities from the posterior sample shown in Fig.\ \ref{hd37124mpdf}.
The points show the observational data and uncertainties.  The lower
panel shows the width of the 68\% credible interval of the posterior
predictive distribution as a function of time.  The width is current
$\simeq5$m s$^{-1}$, but will increase to $\simeq50$m s$^{-1}$ in the
next year.  Observations at these times will be particularly valuable
for refining the orbit of this system.  
\label{hd37124ppdf}}
\end{figure}

\begin{figure}[ht]
\plotone{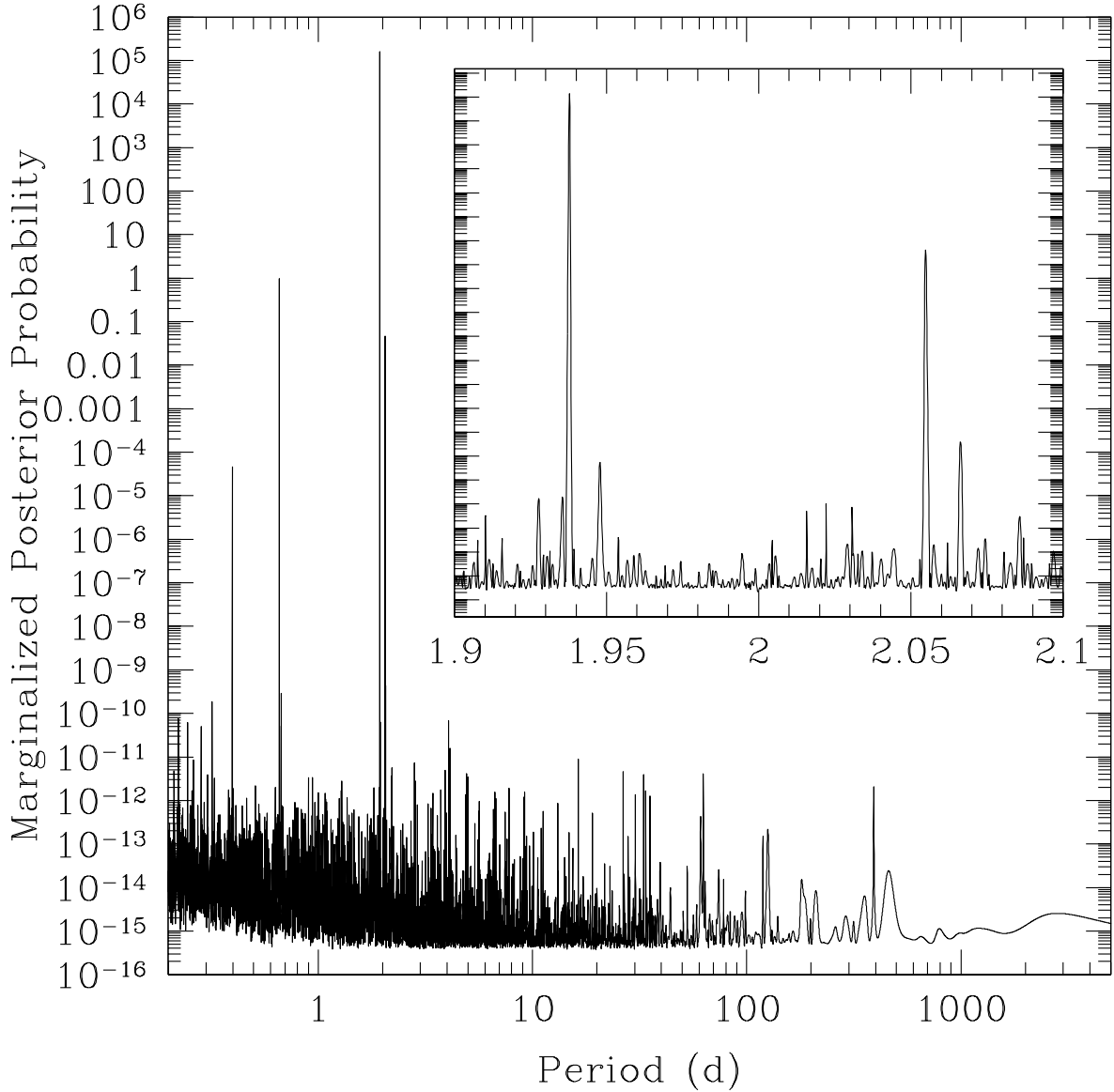} 
\caption[gj876d.eps]{
\noindent
The marginalized posterior probability density as a function
of the the orbital period of the low-mass planet GJ 876 d.
First, we use MCMC to sample from the posterior probability distribution, assuming a model with only two planets on precessing Keplerian orbits.  Then we calculate the posterior predictive residual velocity distributions.  Finally, we calculate the posterior probability density for the orbital period of planet d, which we have assumed to be on a circular orbit.  The inset zooms in on the periodicities near 2 days.  Note that the largest peak contains over 99.9\% of the posterior probability.  
\label{GJ876d}}
\end{figure}

\end{document}